\documentclass[aip,jcp,12pt,showkeys,superscriptaddress,showpacs,preprint,tightenlines,raggedbottom,longbibliography]{revtex4-1}
\usepackage{amsmath}
\usepackage{amsthm}
\usepackage{bm}
\usepackage{graphicx}
\usepackage{hyperref}
\hypersetup{
    colorlinks,
}

\allowdisplaybreaks
\usepackage{wrapfig}
\usepackage{multirow}
\usepackage{nameref}
\usepackage{array}
\usepackage[inline]{enumitem}

\begin{document}
\title{
Investigation of dense manifold of particle-hole excitations in metallic nanowires using r12-correlated frequency-dependent electron-hole interaction kernel
}
\author{Peter F. McLaughlin}
\affiliation
{
Department of Chemistry, Syracuse University, Syracuse, New York 13244 USA
}
\author{Arindam Chakraborty}
\email{archakra@syr.edu}
\affiliation
{
Department of Chemistry, Syracuse University, Syracuse, New York 13244 USA
}
\begin{abstract}
Low-lying electronically excited states in metallic and semiconductor nanoparticles continue to be actively investigated because of their relevance in a wide variety of technological applications. However, first-principles electronic structure calculations on metallic and semiconductor nanoparticles are computationally challenging due to factors such as large system sizes, evaluation and transformation of matrix elements, and high density of particle-hole states. In this work, we present the development of the frequency-dependent explicitly-correlated electron-hole interaction kernel (FD-GSIK) to address the computation bottleneck associated with these calculations. The FD-GSIK method obtains a zeroth-order description of the dense manifold of particle-hole states by constructing a transformed set of dressed particle-holes states. Electron-hole correlation is introduced by using an explicitly correlated, frequency-dependent two-body operator which is local in real-space representation. The resulting electron-hole interaction kernel expressed in an energy-restricted subspace of particle-hole excitations is derived using the Löwdin’s partitioning theory. Finally, the excitation energies are calculated using an iterative solution of the energy-dependent, generalized pseudoeigenvalue equation. The FD-GSIK method was used to investigate low-lying excited states of a series of silver linear clusters and nanowires $(\mathrm{Ag}_n)$. For small clusters, the FD-GSIK results were found to be in good agreement with equation-of-motion cluster-coupled calculations. For nanowires with $n < 50$ , the excitation energy was found to decrease with increasing wire length, and this trend was found to be consistent with EOM-CCSD and time-dependent density functional theory results. However, for $n > 50$ the trend was reversed, and the excitation energy increased with increasing wire length. This trend was found to be consistent with perturbation theory calculations. The results of this investigation demonstrate FD-GSIK is an effective method for investigating electronic excitations and capturing electron-hole correlation in nanomaterials.
\end{abstract}
\maketitle
\section{Introduction} \label{sec:intro}
Noble metal nanowires (NMNWs) such as gold and silver are of great interest for their applications in 
biomedicine,\cite{Haes20046961,Govorov200730,Tallury2010424,Sukirtha2012273,Khlebtsov20101}
catalysis,\cite{Kumar2013658}
energy conversion,\cite{Beck2009,Kang20104378,Xiong20124416,Zengin2015}
and sensing.\cite{West2003285,McFarland20031057,Jain20081578,Feng2012602}
The effect of composition, \cite{LopezLozano20133062,Marinica20121333}
environment, \cite{Malinsky20011471}
size,\cite{Aikens200811272,Kelly2003668,Ross2016816}
and shape \cite{Shabaninezhad2019,Bae201210356,Kelly2003668,Ross2016816}
of noble metal nanoparticles is of particular interest as these cause tunability in the absorption peaks.
In particular, electronically excited states of metal nanowires are of great interest due to the
presence of strong absorption peaks in the visible to near IR region called surface plasmon resonance (SPR). \cite{Eustis2006209,LizMarzan200632}
An SPR is the collective oscillations of electronic transitions due to the density of states within these metallic-like systems. Noble metal nanoparticles have been extensively studied both experimentally\cite{Sonnefraud2012277,Halas20113913,Geisler20171615,Knight20072346} 
and theoretically \cite{Weerawardene2018205,Jain2010153,Varas2016409,Aikens200811272,Morton20113962}
to better understand the plasmonic nature.
\par
\begin{figure}[h!]
\centerline{\includegraphics[scale=1.0]{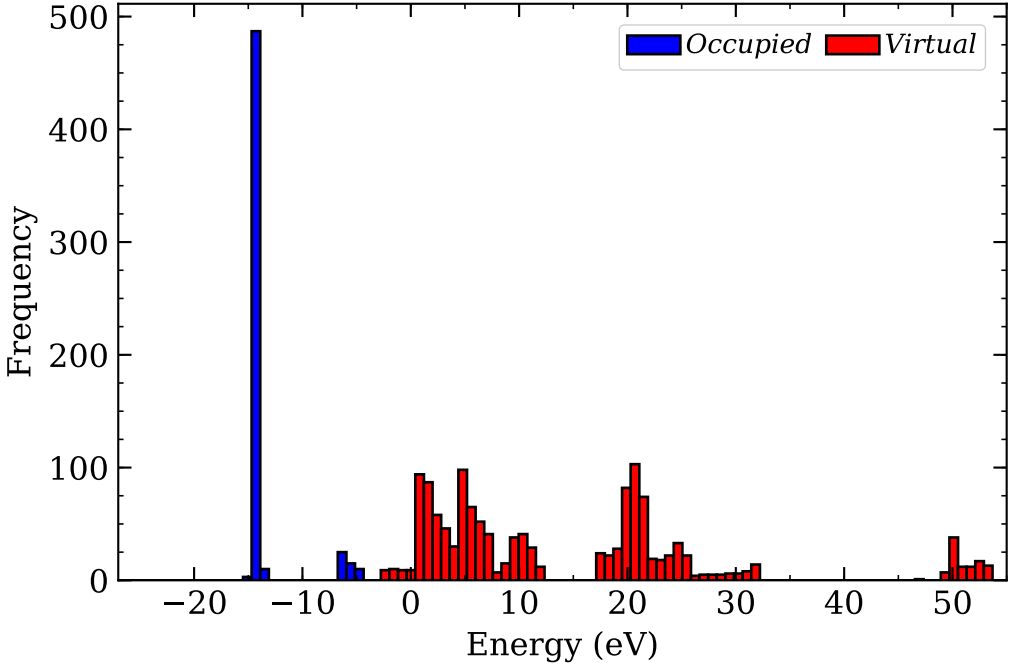}}
\caption{Frequency distribution of molecular orbital energies for Ag$_{100}$ near the HOMO-LUMO gap.}
 \label{fig_Ag_hist}
 \end{figure}

As a consequence of their metallic character, gold and silver nanowires exhibit high density of states
near the band edges.  For example, the distribution of single-particle states obtained from the Hartee-Fock calculation on $\mathrm{Ag}_{100}$ nanowire (\autoref{fig_Ag_hist}) exhibits a high density of states for both occupied and unoccupied orbitals.   
The presence of collective excitation in NMNW is simultaneously the source of their unique photophysical properties and their computational complexity. 
In earlier work, primarily time-dependent density-functional theory (TDDFT) has been used to investigate optical properties on NMNWs. \cite{Bernadotte20131863, Johnson20094445,Gao201413059,Guidez201411512,Conley20195344}

One of the central challenges in performing calculations of NMNWs is the steep scaling of
computational cost with increasing number of atoms in the nanowire.  Most wave function-based-methods require an atomic orbital to molecular orbital (AO-to-MO) transformation of two-electron integrals for post-Hartree-Fock (HF) or post-DFT calculations. 
This transformation is one of the most significant contributors to the overall cost of the calculations and has been focus of research. Naive implementation of the AO-to-MO transformation scales as $O(N^5)$, and various strategies have been developed to reduce the scaling of this transformation. 
Cholesky Decomposition, \cite{Peng20174179,Epifanovsky2013,Koch20039481,Beebe1977683,Krisiloff20155242}
density-fitting, \cite{Krisiloff20155242,Bozkaya2017,Wang20164833,DePrinceIII2014844}
resolution-of-identity, \cite{Epifanovsky2013,Neese20031740}
and the tensor-decomposition techniques\cite{Hohenstein20121085,Parrish2012,Hohenstein2012}
among others have been developed to reduce the cost of this transformation.
In addition to the AO-to-MO transformation, excited-state methods also have to deal with a large space of 
particle-hole excitations for accurate description of the excited state wave function. For example,
in methods such as 
CIS, \cite{Dreuw20054009}
LR-TDDFT, \cite{Dreuw20054009,Casida2012287}
EOM-CCSD, \cite{ShavittBartlett2009}
TD, \cite{Fetter1971}
and GW-BSE, \cite{Onida2002601,Blase20181022,Govoni20152680}
the excited state calculations
need construction of the $\mathbf{A}$ and $\mathbf{B}$ response matrices in the one-particle one-hole (1p-1h) basis. The cost of excited state calculations is exacerbated for multireference wave functions.
For TDDFT methods, the real-time formulation (RT-TDDFT) as opposed to the linear-response (LR-TDDFT)
provide an efficient alternate route for a computationally efficient procedure to calculate excited state properties. A recent review by Weerawawardene and Aikens provides a detailed
comparison of these TDDFT methods for noble metal nanoparticles.\cite{Weerawardene201827}
The real-time propagation approach allows for a direct treatment of a collection of 1p-1h states and is better suited for treating a dense manifold of particle-hole excitations. 
Recently, the efficacy of this approach has been demonstrated in a series of studies on Ag\cite{Ding2014,Peng20156421} and Au nanowires.\cite{Gao2012,Senanayake201914734} 
Time-dependent density-functional theory has demonstrated impactful insight into understanding the plasmon resonance within 
many NMNW and nanoparticles in obtaining electron-correlation.\cite{Piccini201317196,Barcaro201412450,Baseggio201612773,Gao201413059,
Ma201617044,Zhang2014635,Fernando20156112,Senanayake201914734}
Although TDDFT-based methods have proved useful, limited work has been done with other excited state methods such as
configuration interaction (CI), \cite{Guidez201415501,Bae20128260}
complete active space self-consistent field (CASSCF) \cite{Fales20174162},
and equation-of-motion coupled-cluster (EOM-CC) \cite{Koutecky200110450}
 that work towards understanding the multireference characteristics of the collective excitations within these systems, which has resulted in restricting the investigations to a few atoms or active space,  due to the dense manifold of single-particle states and cost of the overall computation for these noble metal nanoparticles.
However, both wave function and RT-TDDFT implementation require AO-to-MO transformations of integrals which add to the overall computational cost for these systems. 
\par
In this work, we present a first-principle real-space wave function based approach which avoids AO-to-MO integral transformation by calculating all the necessary integrals directly in the MO basis using the Monte Carlo techniques. 
The method, uses an explicitly correlated frequency-dependent electron-hole interaction kernel (FD-GSIK) 
for including electron-hole correlation in calculation of excited states in many-electron systems.  \cite{doi:10.1021/acs.jctc.9b01238} 
The treatment of dense manifold of particle-hole states in FD-GSIK method is achieved by introducing 
dressed field operators $\{\hat{\Psi}(\mathbf{r}),\hat{\Psi}^\dagger(\mathbf{r})\}$ that represent collective quasiparticle coordinate. Inspired by the work by Li and co-workers on energy-specific 
TDDFT \cite{Goings2016,Liang20113540}
 and 
 EOM-CCSD \cite{Peng20154146}
 techniques of excited state calculations, an energy-based partitioning scheme is used to define the quasiparticle field operators. 
The combination of a real-space formulation using partitioned field operators with explicitly correlated 
frequency dependent kernel allows us to overcome the computational barriers mentioned above.
We have applied the developed method for studying low-lying excited states of a series silver nanowires consisting of $[\mathrm{Ag}_2,\dots,\mathrm{Ag}_{100}]$ and have benchmarked them against EOM-CCSD and linear response TDDFT calculations. The theoretical details of the derivation, its computational implementation, and its application to metallic nanowires are presented in the following sections.  

\section{Theory} \label{sec:theory}
\subsection{Energy-restricted quasiparticle creation operators}
We start by defining the quasiparticle field operators $\hat{\Psi}_\mathrm{h}$ and $\hat{\Psi}_\mathrm{e}$. The field operators, defined using the set of one-particle states, are obtained from the eigenspectum of the Fock operator
\begin{align}
	f \chi_p(\mathbf{x}) = \epsilon_p \chi_p(\mathbf{x}).
\end{align}
Using the MOs $\chi_p$, the quasi-electron and quasi-hole field operators are expressed as
\begin{align}
	\hat{\Psi}_\mathrm{h}^\dagger (\mathbf{x}) = \sum_p \theta(\epsilon_\mathrm{HOMO}-\epsilon_p) \chi_p^\ast(\mathbf{x}) p^\dagger,
\end{align}
\begin{align}
	\hat{\Psi}_\mathrm{e}^\dagger (\mathbf{x}) = \sum_p \theta(\epsilon_p-\epsilon_\mathrm{LUMO}) \chi_p^\ast(\mathbf{x}) p^\dagger,
\end{align}
where $\epsilon_\mathrm{HOMO}$  is the energy of the highest occupied molecular orbital (HOMO),  $\epsilon_\mathrm{LUMO}$  is the energy of the lowest unoccupied molecular orbital (LUMO), and $\theta$ is the Heaviside function.
The effective many-body electron-hole Hamiltonian has the general form \cite{Zhu199613575,Woggon1996,Mattuck1976,Elward2012,Ellis2016188,Ellis20171291}
\begin{align}
	H_\mathrm{eh} = H^0_\mathrm{eh} + V_\mathrm{ee} + 
	V_\mathrm{hh} + V_\mathrm{eh},
\end{align}
where
\begin{align}
	H^0_\mathrm{eh} 
   &= 
   \int d\mathbf{x} \hat{\Psi}_\mathrm{h}^\dagger(\mathbf{x}) f \hat{\Psi}_\mathrm{h}(\mathbf{x})
   +  
   \int d\mathbf{x} \hat{\Psi}_\mathrm{e}^\dagger(\mathbf{x}) f \hat{\Psi}_\mathrm{e}(\mathbf{x}),
\end{align}
\begin{align}
	V_\mathrm{ee}
	&= 
	\int d\mathbf{x} d\mathbf{x}'
	\hat{\Psi}_\mathrm{e}^\dagger(\mathbf{x})
	\hat{\Psi}_\mathrm{e}^\dagger(\mathbf{x}')
	 w_\mathrm{ee}(\mathbf{x},\mathbf{x}') 
	 \hat{\Psi}_\mathrm{e}(\mathbf{x}')
	 \hat{\Psi}_\mathrm{e}(\mathbf{x}),
\end{align}
\begin{align}
	V_\mathrm{hh}
	&= 
	\int d\mathbf{x} d\mathbf{x}'
	\hat{\Psi}_\mathrm{h}^\dagger(\mathbf{x})
	\hat{\Psi}_\mathrm{h}^\dagger(\mathbf{x}')
	 w_\mathrm{hh}(\mathbf{x},\mathbf{x}') 
	 \hat{\Psi}_\mathrm{h}(\mathbf{x}')
	 \hat{\Psi}_\mathrm{h}(\mathbf{x}),
\end{align}
\begin{align}
	V_\mathrm{eh}
	&= 
	\int d\mathbf{x} d\mathbf{x}'
	\hat{\Psi}_\mathrm{e}^\dagger(\mathbf{x})
	\hat{\Psi}_\mathrm{h}^\dagger(\mathbf{x}')
	 w_\mathrm{eh}(\mathbf{x},\mathbf{x}') 
	 \hat{\Psi}_\mathrm{h}(\mathbf{x}')
	 \hat{\Psi}_\mathrm{e}(\mathbf{x}),
\end{align}
and $w_\mathrm{ee}$, $w_\mathrm{hh}$, and $w_\mathrm{eh}$ are the
quasiparticle interaction operators.   
The electron-hole Hamiltonian can be factored into a sum of noninteraction terms and interaction terms and 
 for a 1-particle 1-hole system the total Hamiltonian can be expressed as
\begin{align}
	H_\mathrm{eh} &= H^0_\mathrm{eh} + V_\mathrm{eh}.
\end{align}
In addition to the quasiparticle field operators, we also defined a set of energy-restricted hole and particle creation operators$\{D^\dagger_\mathrm{h},D^\dagger_\mathrm{e}\}$,
\begin{align}
\label{eq:eta}
	D_\mathrm{h}^\dagger(\eta_\mathrm{h})
	&= \sum_{i}^{N_\mathrm{occ}}
	\theta(\epsilon_i - \eta_h)
	 i^\dagger,\\
	D_\mathrm{e}^\dagger(\eta_\mathrm{e})
	&= \sum_{a}^{N_\mathrm{vir}}
	\theta(\eta_\mathrm{e}-\epsilon_a)
	 a^\dagger,
\end{align}
where $\eta_\mathrm{h}$ and $\eta_\mathrm{e}$ are the energy cutoff parameters. 
Using these operators, we define the following particle-hole state $\vert P \rangle$,
\begin{align}
	\vert P \rangle
	= 
	\sqrt{\frac{1}{M_P}}
	D_\mathrm{h}^\dagger(\eta_\mathrm{h})
	D_\mathrm{e}^\dagger(\eta_\mathrm{e})
	\vert 0_\mathrm{h}  0_\mathrm{e}\rangle, 
\end{align}
where $M_P$ is the normalization constant ensuring $\langle P \vert P \rangle = 1$.
The goal of this work is to derive the many-body correction to the collection of 1p-1h excitations 
represented by the state vector $\vert P \rangle$.  
\par
The state $\vert P \rangle$ represents the zeroth-order description to the electron-hole wavefunction
\begin{align}
	\vert \Psi_X\rangle^{(0)} = \vert P \rangle,
\end{align} 
and the zeroth-order excitation energy $\omega_P^0$ is calculated from the following expectation value.
\begin{align}
\label{eq:omega0}
	\omega_P^0 
	= \langle P \vert H_\mathrm{eh}^0 \vert P \rangle.
\end{align}
The first-order correction to $\omega_P^0$ is given by 
\begin{align}
	\omega_P^{(1)} = \omega_P^0 + \langle P \vert r_{12}^{-1} \vert P \rangle_A,
\end{align}
where the subscript A in $\langle \dots \rangle_A$ represents an antisymmetrized matrix element.
To go beyond the first-order approximate, a correlated description of the electron-hole wave function is needed that includes contributions from states that are orthogonal to $\vert P\rangle$.
The general form of such a correlated electron-hole wavefunction $\Psi_X$ can be written as,
\begin{align}
	\vert \Psi_X\rangle
	&= C_{P} \vert P \rangle + C_\perp \vert P_\perp \rangle,
\end{align}
where $C_{P}$ and $C_\perp$ are expansion coefficients and $P_\perp$ 
is the correlated wavefunction that exists in the orthogonal subspace of $P$.
This is a very generic form and both many-body perturbation theory (MBPT) and configuration interaction (CI) wavefunctions can be 
expressed in this form, where choice and construction of $P_\perp$ differentiates between 
treatment of electron-hole correlation within different methods. In the present work, 
we use a real-space explicitly-correlated operator approach and use the electron-hole correlator operator 
$\Lambda$ to express $P_\perp$ as,\cite{doi:10.1021/acs.jctc.9b01238}
\begin{align}
	\vert \Psi_X\rangle
	&= C_{P} \vert P \rangle + C_{Q} \Lambda(\omega,r_\mathrm{eh}) \vert Q \rangle,
\end{align}
where  $\vert Q \rangle$ spans the orthogonal subspace of $P$. The expansion coefficients and the operator $\Lambda$ are determined by a projective solution of the electron-hole Hamiltonian
\begin{align}
\label{eq:hpsi}
H_{\mathrm{eh}}\vert \Psi_X\rangle = \omega \vert \Psi_X\rangle
\end{align}

\subsection{Construction of frequency-dependent electron-hole interaction kernel}
The construction of the frequency-dependent geminal-screened electron-hole interaction kernel (FD-GSIK) has been derived earlier and a brief summary of the key steps are presented below.\cite{doi:10.1021/acs.jctc.9b01238}
In the first step, the eigenvalue equation \autoref{eq:hpsi} is projected 
onto the   $\langle P \vert$ and  $\langle Q \vert$ subspace to obtain
the following matrix equation,
\begin{align}
\begin{bmatrix}
 \langle P \vert H_\mathrm{eh} \vert P \rangle 
  & 
 \langle P \vert H_\mathrm{eh} \Lambda(\omega) \vert Q \rangle \\
 \langle Q \vert H_\mathrm{eh} \vert P \rangle 
 & 
 \langle Q \vert H_\mathrm{eh} \Lambda(\omega) \vert Q \rangle
\end{bmatrix}
\begin{bmatrix}
 C_{P} \\
 C_{Q}
\end{bmatrix}
&= \omega
\begin{bmatrix}
 1   &  \langle P \vert  \Lambda(\omega) \vert Q \rangle\\
 0  & 
 \langle Q \vert  \Lambda(\omega) \vert Q \rangle
\end{bmatrix}
\begin{bmatrix}
 C_{P} \\
 C_{Q}
\end{bmatrix}.
\end{align}
In the second step, L{\"{o}}wdin's partitioning is performed to derive the electron-hole interaction kernel,
\begin{align}
	\langle P \vert H_\mathrm{eh} \vert P \rangle  C_P
	+
	\langle P \vert H_\mathrm{eh} \Lambda(\omega) \vert Q \rangle C_Q
	&=
	\omega C_P
	+ 
	\langle P \vert \Lambda(\omega) \vert Q \rangle C_Q \\
	\langle Q \vert H_\mathrm{eh} \vert P \rangle C_P
    +
    \langle Q \vert H_\mathrm{eh} \Lambda(\omega) \vert Q \rangle C_Q
	&=
	\omega \langle Q \vert H_\mathrm{eh} \Lambda(\omega) \vert Q \rangle C_Q,
\end{align}
Rearranging the equations,
\begin{align}
	\langle P \vert H_\mathrm{eh} \vert P \rangle  C_P
	+
	\left[
	\langle P \vert H_\mathrm{eh} \Lambda(\omega) \vert Q \rangle 
	-
	\langle P \vert \Lambda(\omega) \vert Q \rangle
	\right] 	 C_Q
	&=
	\omega C_P \\
	\left[
	\omega \langle Q \vert H_\mathrm{eh} \Lambda(\omega) \vert Q \rangle 
   -\langle Q \vert H_\mathrm{eh} \Lambda(\omega) \vert Q \rangle
	\right]   
    C_Q
    &=
    \langle Q \vert H_\mathrm{eh} \vert P \rangle C_P,
\end{align}
and eliminating $C_Q$ gives the equation for the electron-hole interaction kernel $K_{PP}(\omega)$
\begin{align}
\label{eq_FDGSIK_iterative}
\left[ H_{PP} + K_{PP}(\omega) \right] C_{P} = \omega C_{P},
\end{align}
The closed-form analytical expression for the electron-hole interaction kernel, $K_{PP}(\omega)$ is given by
\begin{align}
\label{eq_Kpp}
	K_{PP}(\omega)
	&=
	-\frac{
		\left[
		\langle P \vert r_{12}^{-1} \Lambda(\omega) \vert Q \rangle_A
	    - (\omega_{P}^0 - \omega)\langle P \vert \Lambda(\omega) \vert Q \rangle
		\right]
	\langle Q \vert r_{12}^{-1} \vert P \rangle_A 
	} 
	{  
 	 \langle Q \vert r_{12}^{-1} \Lambda(\omega) \vert Q \rangle_A 
 	 + (\omega_{Q}^0  -  \omega ) \langle Q \vert \Lambda(\omega) \vert Q \rangle
    }
\end{align}
where $\langle \dots \rangle_A$ represents antisymmetrized matrix elements.
\autoref{eq_Kpp} is a non-linear equation and is solved iteratively starting with $\omega_P^0$ as the first guess. 
The form of the electron-hole correlator operator $\Lambda(\omega,r_\mathrm{eh})$ is given by
\begin{align} 
\label{eq_lambda}
	\Lambda(\omega,r_\mathrm{eh})
	&= \frac{(\omega_{P}^0 - \omega)}{\gamma} 
	e^{-r_\mathrm{eh} \gamma }
\end{align}
and is an explicitly-correlated frequency-dependent real-space operator. The derivation of this operator
has been presented earlier\cite{doi:10.1021/acs.jctc.9b01238}, and is related to infinite-order partial summation to particle-hole diagrams.\cite{Bayne20183656}  
Other than $\omega$, which is obtained from the iterative solution of \autoref{eq_Kpp}, the operator depends on two additional parameters, $\omega_P^0$ and $\gamma$, both of which are evaluated during the course of the calculation. Parameter $\omega_P^0$ was defined earlier in \autoref{eq:omega0} and $\gamma$
is defined as
\begin{align}
\label{eq:gamma}
	\gamma = \langle P \vert \frac{1}{r_\mathrm{eh}}\vert P \rangle.
\end{align}
Computer implementation and iterative solution of \autoref{eq_Kpp} require two additional components. 
The first is the definition and construction of the orthogonal function $Q$ and the second is the evaluation of the matrix elements and both of these steps are presented in \autoref{sec_opt_dQ} and \autoref{sec:mc}, respectively.
\subsection{Construction of the $\vert Q \rangle$ state} \label{sec_opt_dQ}
\label{sec:q}
 The basis vector $\vert Q \rangle$ is constructed from the direct product of
 particle and hole states that are not included in the $\vert P \rangle$ state, 
\begin{align}
\label{eq_dressed_Q}
\vert Q \rangle &=\vert h' \rangle \otimes \vert e' \rangle
\end{align}
where $\vert h' \rangle$ and $\vert e' \rangle$ can be viewed as dressed 
particle and hole states,
\begin{align}
\vert h' \rangle &= \sum_{j \notin P} c_j^\mathrm{h} \vert j \rangle, \\
\vert e' \rangle &= \sum_{b \notin P} c_b^\mathrm{e} \vert b \rangle,
\end{align}
obtained from a linear combination of bare particle and hole states not included in $\vert P \rangle$.
The choice of the expansion coefficient directly impacts the form of $Q$ and the construction of the electron-hole interaction kernel. A natural choice for determination of the 
expansion coefficients are by minimization of the trace of the electron-hole Hamiltonian 
\begin{align}
	\min_{\mathbf{c}^\mathrm{h},\mathbf{c}^\mathrm{e}}
	\left[
	\langle P \vert H_\mathrm{eh} \vert P \rangle + \langle Q \vert H_\mathrm{eh} \vert Q \rangle
	\right] 
    \rightarrow  	
	\left\{
	\mathbf{c}^\mathrm{h}_\mathrm{opt},\mathbf{c}^\mathrm{e}_\mathrm{opt}
	\right\}.
\end{align}
Since $P$ is independent of the expansion coefficients, the above minimization reduces to
\begin{align}
	\min_{\mathbf{c}^\mathrm{h},\mathbf{c}^\mathrm{e}}
	\left[
	\omega_Q^0 + \langle Q \vert V_\mathrm{eh} \vert Q \rangle
	\right] 
    \rightarrow  	
	\left\{
	\mathbf{c}^\mathrm{h}_\mathrm{opt},\mathbf{c}^\mathrm{e}_\mathrm{opt}
	\right\},
\end{align}
with the following normalization constraints
\begin{align}
	\mathbf{c}^{\mathrm{h}\dagger} \mathbf{c}^\mathrm{h}  &= 1, \\
	\mathbf{c}^{\mathrm{e}\dagger} \mathbf{c}^\mathrm{e}  &= 1.
\end{align}
Although the above procedure will generate an optimized state vector Q that is suitable for 
performing L{\"{o}}wdin's partitioning, the procedure is computationally demanding because it 
requires calculation of a large number of particle-hole $V_\mathrm{eh}$ elements. 
For this reason, we add additional restrictions to the form of the expansion coefficient.
First, we make the expansion coefficients to be proportional to the molecular orbital energies,
\begin{align}
	c_j^\mathrm{h} \propto e^{-\alpha^\mathrm{h} (\epsilon_j - \epsilon_\mathrm{Qmin}^\mathrm{h}) }, \\
	c_b^\mathrm{e} \propto e^{-\alpha^\mathrm{e} (\epsilon_b - \epsilon_\mathrm{Qmin}^\mathrm{e}) },
\end{align}
where $\epsilon_\mathrm{Qmin}^\mathrm{h}$ and $\epsilon_\mathrm{Qmin}^\mathrm{e}$ are the lowest hole and particle energies in state $\vert Q \rangle$,
\begin{align}
	\epsilon_\mathrm{Qmin}^\mathrm{h} = \min_{ j \in Q } \epsilon_j, \\
	\epsilon_\mathrm{Qmin}^\mathrm{e} = \min_{ b \in Q } \epsilon_b.
\end{align}
Using the above expression, the multi-parameter optimization reduces to a two-parameter optimization,
\begin{align}
	\min_{\alpha^\mathrm{h},\alpha^\mathrm{e}}
	\left[
	\omega_Q^0 + \langle Q \vert V_\mathrm{eh} \vert Q \rangle
	\right] 
    \rightarrow  	
	\left\{
	\mathbf{c}^\mathrm{h}_\mathrm{opt},\mathbf{c}^\mathrm{e}_\mathrm{opt}
	\right\}.
\end{align}
In the second step, we replace the exact integral $ \langle Q \vert V_\mathrm{eh} \vert Q \rangle$
by an approximate integral which is based on the 2-particle density. 
The potential energy term can be expressed in terms of density
\begin{align}
\langle{Q} \vert V_\mathrm{eh} \vert {Q} \rangle &= \langle \rho_{Q} r_{\mathrm{eh}}^{-1} \rangle,
\end{align}
where the density $ \rho_{Q} $ is defined as
\begin{align}
 \rho_{Q}(1,2) &= Q^\ast(1,2)Q(1,2) \\
 									&= {h}'(1){e}'(2){h}'(1){e}'(2) \\
 									&= {h}'(1){h}'(1){e}'(2){e}'(2) \\
 \rho_{Q}(1,2) &=  \rho_{{h}'}(1) \rho_{{e}'}(2).
\end{align}
The particle and hole densities can be calculated from the MO densities,
\begin{align}
\rho_{{h}'}(\mathbf{r}) &= \sum_{j \in \{ Q \}} c_{j}^2 \rho_j(\mathbf{r}), \\
\rho_{{e}'}(\mathbf{r}) &= \sum_{b \in \{ Q \}} c_{b}^2 \rho_b(\mathbf{r}).
\end{align}
We reduce the computational cost by approximating the MO densities by their Gaussian approximation,
\begin{align}
	\rho_j(\mathbf{r})
	&\approx \rho_j^{G}(\mathbf{r};\mu_j,\sigma_j) \\
	\rho_b(\mathbf{r})
	&\approx \rho_a^{G}(\mathbf{r};\mu_b,\sigma_b)
\end{align}
where $\rho^{G}(\mathbf{r})$ is the isotropic 3D normal probability distribution function
\begin{align}
	\rho^{G}(\mathbf{r};\mu,\sigma)
	&=
	\left(\frac{1}{\sqrt{2\pi \sigma^2}}\right)^3
	e^{-\frac{1}{2}\frac{(x-\mu)^2}{\sigma^2}}
	e^{-\frac{1}{2}\frac{(y-\mu)^2}{\sigma^2}}
	e^{-\frac{1}{2}\frac{(z-\mu)^2}{\sigma^2}}
\end{align}
The defining coefficients for the Gaussian densities $(\mu,\sigma)$ are obtained 
by minimizing the Kullback\textendash Leibler divergence\cite{Kullback1968} between the true MO densities $ \rho_j,\rho_b$ and the Gaussian probability distribution functions
\begin{align}
	D_\mathrm{Kullback-Leibler}
	&=
	\int_V d\mathbf{r} \rho_{j}
	\ln\left[ 
	\frac{\rho_j}{\rho_j^{G}}	
	\right].
\end{align}
It is important to note that the approximate Gaussian densities derived above are only used for determination of the expansion coefficients and not in the eigenvalue equation to determine $\omega$. All matrix elements needed to construct the electron-hole interaction kernel are obtained using Monte Carlo integration and is described in the following section. 

\subsection{Monte Carlo integration}
\label{sec:mc}
One of the advantages of the present method is that it circumvents the AO-to-MO transformation of the 
two-electron integrals. In the FD-GSIK method, the integrals over MO are computed directly in real-space. 
For any point $\mathbf{r}$, the value of a general spatial molecular orbital $\psi_p(\mathbf{r})$ is evaluated as 
\begin{align}
	\psi_p(\mathbf{r})
	&=
	\sum_{\mu}^{N_\mathrm{AO}}
	C_{\mu p} \phi_\mu(\mathbf{r}),
\end{align}
where $C_{\mu p}$ are MO eigenvectors, $\phi_\mu(\mathbf{r})$  are the atomic orbitals, 
and $N_\mathrm{AO}$ is the total number of AOs used in the calculation. The FD-GSIK method requires evaluation of two-body integrals with the general form of
\begin{align}
\label{eq:Ipq}
	I_{pq,p'q'}(\Omega)
	= 
	\int_{-\infty}^{+\infty} d\mathbf{r} d\mathbf{r'}
	\psi_p(\mathbf{r}) \psi_q(\mathbf{r}) 
	\Omega(\mathbf{r},\mathbf{r}')
	\psi_{p'}(\mathbf{r}') \psi_{q'}(\mathbf{r}') 
	=	
	\int_{-\infty}^{+\infty} d\mathbf{r} d\mathbf{r'}
	K_{pq;p'q'}(\mathbf{r},\mathbf{r}'),
\end{align}
where the two-body kernel can be
\begin{align}
	\Omega(\mathbf{r},\mathbf{r}')
	&= 
	\vert \mathbf{r}-\mathbf{r}' \vert^{-1},
	\exp\left[ -\gamma
	\vert \mathbf{r}-\mathbf{r}' \vert^{-1}
	\right]
	,
	\vert \mathbf{r}-\mathbf{r}' \vert^{-1} \exp\left[ -\gamma
	\vert \mathbf{r}-\mathbf{r}' \vert^{-1}
	\right].
\end{align}
To evaluate the two-body integral in \autoref{eq:Ipq} stochastically, we
also define the following reference integral whose solution is known analytically,
\begin{align}
\label{eq:I0pq}
	I_{pq,p'q'}^0
	= 
	\int_{-\infty}^{+\infty} d\mathbf{r} d\mathbf{r'}
	\left(
	\frac{\psi_p^2 (\mathbf{r})+\psi_q^2(\mathbf{r})}{2}
	\right)
	\left(
	\frac{\psi_{p'}^2 (\mathbf{r}')+\psi_{q'}^2(\mathbf{r}')}{2}
	\right)
	=\int_{-\infty}^{+\infty} d\mathbf{r} d\mathbf{r'}
	K_{pq;p'q'}^0(\mathbf{r},\mathbf{r}')
	= 1.
\end{align}
Using the above reference integral $I_{pq,p'q'}^0$, the kernel integral $I_{pq,p'q'}(\Omega)$ is expressed
as
\begin{align}
	I_{pq,p'q'}(\Omega) 
	=
	\frac{I_{pq,p'q'}(\Omega) }{I_{pq,p'q'}^0} 
	\approx
	\frac{\sum_{\mathbf{r},\mathbf{r}' \in \mathcal{S}_\mathrm{MC} } K_{pq;p'q'}(\mathbf{r},\mathbf{r}') }
	{\sum_{\mathbf{r},\mathbf{r}' \in \mathcal{S}_\mathrm{MC} } K_{pq;p'q'}^0(\mathbf{r},\mathbf{r}')}
\end{align}
where $\mathcal{S}_\mathrm{MC}$ is the set of sampling points and $K_{pq;p'q'}^0$, $K_{pq;p'q'}$
are the kernels defined in \autoref{eq:I0pq} and \autoref{eq:Ipq}, respectively.
A simple uniform sampling leads to very slow convergence of the above integral with respect
to increasing number of sampling points and leads to very inefficient Monte Carlo calculation. 
This is a well-known issue\cite{kalos2009monte} and strategies such as importance sampling and stratified sampling have been developed to increase the convergence and accuracy of Monte Carlo calculations.\cite{kalos2009monte}
In this work, we used the composite control-variate stratified sampling (CCVSS) approach which was specifically developed for efficient calculation of molecular integrals.\cite{bayne2018development}

\section{Results}\label{sec:results}
\subsection{Chemical systems and computational details}
The FD-GSIK method with the dressed quasiparticle orbitals was used for investigating
low-energy electronic excited states of a series of linear silver clusters and nanowires with the stoichiometry of $\mathrm{Ag}_2 \dots \mathrm{Ag}_{100}$. The nanowires were constructed using a bond distance of 3.51 \r{A}, which was obtained by using the central Ag-Ag bond distance obtained from an HF-optimized $\mathrm{Ag}_{100}$ calculation with LANL2DZ basis and effective core potential (ECP).
The uncorrelated ground state single-particle states were calculated using the HF method and LANL2DZ basis and ECP for the nanowires with the TERACHEM package.\cite{Seritan2020224110}
The Q-Chem package \cite{QCHEM_shortlist} was used to obtain EOM-CCSD excitation energies for a subset of the  Ag$_{n}$ nanowires using LANL2DZ basis and ECP.
\par
The spatial MO $\psi(\mathbf{r})$ grids were evaluated with 100 points per dimension with grid boundaries that were set by a cutoff tolerance of $\vert \psi(\mathbf{r}) \vert \leq 10^{-12}$.
This resulted in a $10^6$ point dense grid that was used for construction of the dressed MOs.
The grid density was verified by calculating the normalization integral for all MOs in the system. The point density for the grid was optimized for the largest system ($\mathrm{Ag}_{100}$) and was subsequently used for all the remaining systems in the set. Monte Carlo numerical integration was performed for evaluation of all necessary matrix elements described in \autoref{sec:theory} and a total of $10^{10}$ sampling points was used for each integral. 
\par
We performed two different sets of calculations utilizing different energy cutoff parameters 
$(\eta_h,\eta_e)$ for construction of the $\vert P \rangle$ state using \autoref{eq:eta}.
The first set of calculations presented in \autoref{sec:homo} focused on only on the lowest energy excitation
and the parameter were set to $\eta_h = \epsilon_{HOMO}$ and  $\eta_e = \epsilon_{LUMO}$. 
The second set of calculations presented in \autoref{sec:collective}, investigated $\vert P \rangle$ that included a collection of particle-hole excitations that originate from the dense manifold of particle and hole states near the HOMO and LUMO states. Specifically, the energy cutoff parameters 
$(\eta_h,\eta_e)$ were selected such that $\omega_P^0$ included all particle-hole excitations within 1eV of 
the HOMO-LUMO excitation.  The values of $(\eta_h,\eta_e)$ and number of particle and hole orbitals included in $\vert P \rangle$ are presented in \autoref{tab:dressed_Ag_details} of Appendix A.
\subsection{Particle-hole excitation from discrete states}
\label{sec:homo}
\begin{figure}[h!]
\centerline{ \includegraphics[scale=1.0]{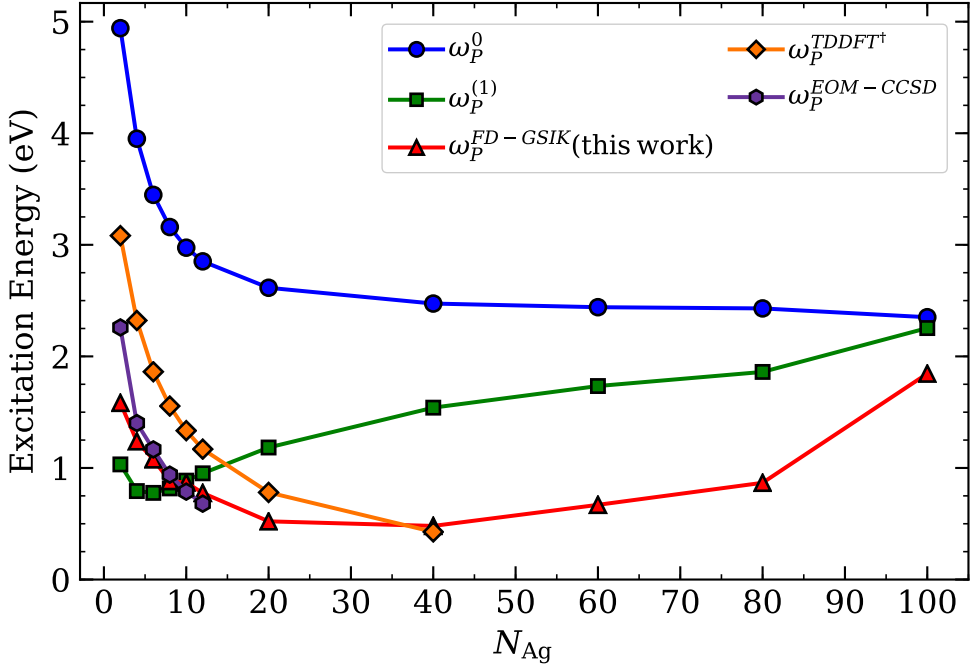}}
\caption{Excitation energy of Ag nanowires using FD-GSIK for the HOMO-LUMO gap in comparison to reported TD-DFT results$\dagger$\cite{Guidez20124190} and EOM-CCSD.}
\label{fig_Ag_HL}
 \end{figure}

The results for particle-hole excitation between the HOMO and LUMO states are presented in \autoref{fig_Ag_HL}. 
The calculations showed that the zeroth-order excitation energies $\omega_P^0$ obtained from the HF
calculations decreased monotonically with increasing length. 
The variance in $\omega_P^0$ also decreased with increasing length. 
In contrast to the HF results, the FD-GSIK curve exhibited a point of inflection. 
Specifically, for smaller wire lengths $n \le 40$, $\omega^\mathrm{FD-GSIK}$ initially decreases but then changes slope and increases with increasing length. 
Analogous to the FD-GSIK results, the results from the first-order perturbation theory $\omega^{(1)}$, also exhibited similar scaling behavior with increasing number of atoms. 
The FD-GSIK results were also compared to EOM-CCSD calculations and previously reported linear response TDDFT calculations by Guidez and Aikens.\cite{Guidez20124190}
Both EOM-CCSD and LR-TDDFT results also showed decreased excitation energy with increasing number of atoms for $n \le 20$. These observations were consistent with the results from first-order PT and the FD-GSIK results. 
 
\subsection{Particle-hole excitation from dense manifold of states}
\label{sec:collective}
\begin{figure}[h!]
\centerline{\includegraphics[scale=1.0]{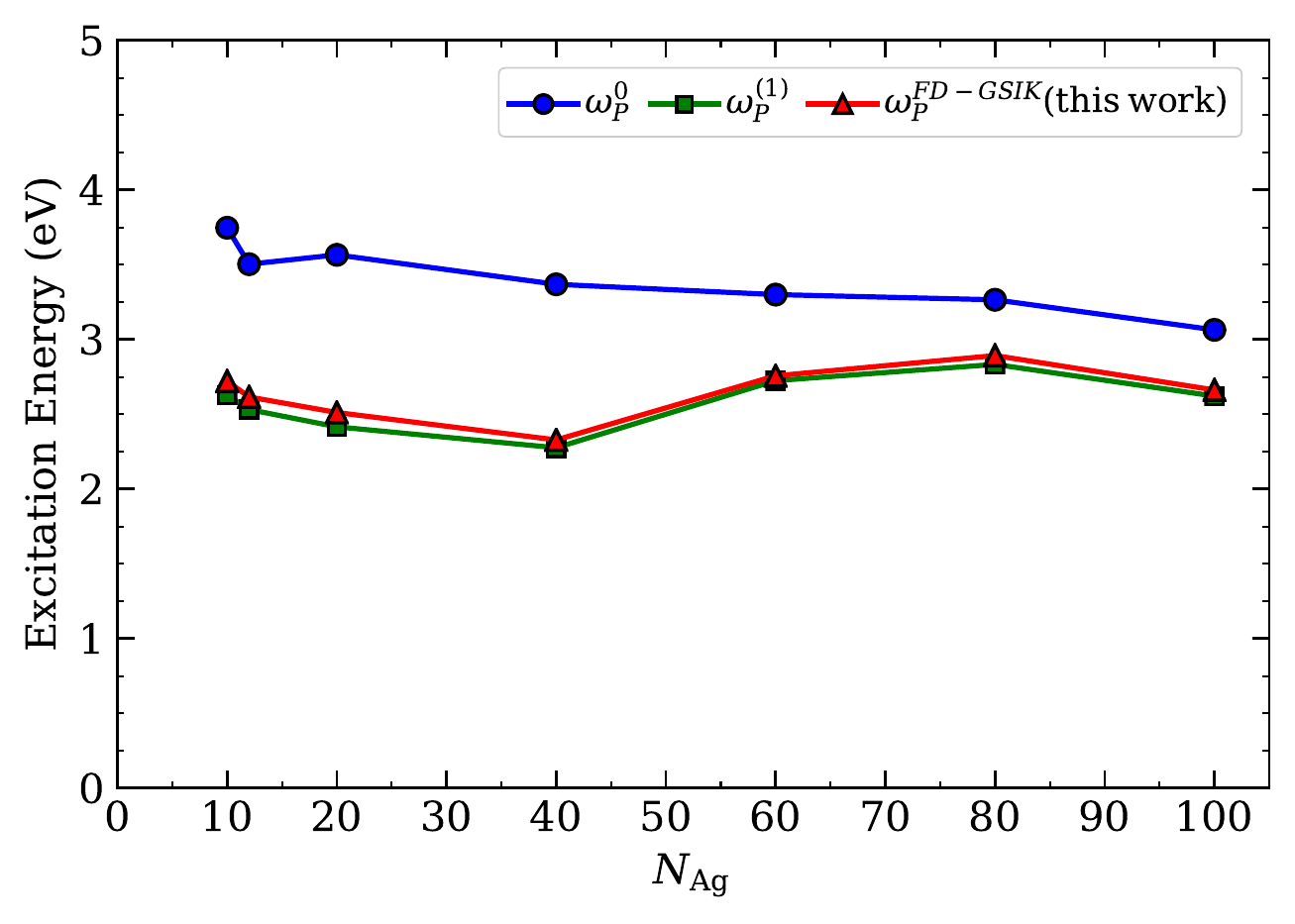}}
\caption{Comparison of the uncorrelated gap to the first-order corrected and FD-GSIK correlated excitation obtained for $\vert P \rangle$ in Ag nanowires.}
\label{fig_Ag_dressed}
 \end{figure}

We investigated electronic excitation from a dense manifold of particle and hole states.
For example, as shown in \autoref{fig_Ag_hist}, $\mathrm{Ag}_{100}$ exhibit high density of states near the HOMO-LUMO gap. In this investigation, all particle-hole excitations that are within 1 eV of the HOMO-LUMO gap were included in the construction of the $\vert P \rangle$ state.
For example, the  Ag$_{100}$ nanowire calculations included 17 hole ($N_{P^h}$) and 14 particle  ($N_{P^e}$) orbitals for construction of $\vert P \rangle$. The energy window was selected by choosing the appropriate values of the $\eta_\mathrm{h}$ and $\eta_\mathrm{e}$ and are presented in \autoref{tab:dressed_Ag_details}. The calculations were performed for a series of silver nanowires and the results are presented in \autoref{fig_Ag_dressed}. 
For the HF frequencies $\omega_P^0$ an overall red-shift was observed with respect to increasing chain length.  
The FD-GSIK results were more complex and, unlike the HF calculations, exhibited non-monotonic behavior with respect to increasing chain length.  Initially the FD-GSIK excitation energy was found to decrease until
$N=50$, after which it started increasing again. Interesting, a similar trend was also exhibited by excitation energy calculated from first-order perturbation theory. The first-order PT results show surprisingly good agreement with the FD-SIK results. The results also indicate that periodic-boundary calculations on an infinitely long nanowire should be performed, and future work on this system will focus on the periodic-boundary implementation of the FD-GSIK method. 

\subsection{Error analysis and timing data}
In the FD-GSIK formulation, the use of a Monte Carlo integration scheme
allows us to reduce the numerical error in the integrals systematically. 
Presented in \autoref{tab:homo_lumo_Ag_std} and \autoref{tab:dressed_Ag_std} are the standard deviations of the FD-GSIK excitation energies for all systems under investigation.
The timing data is presented for the $1+10$ MC runs, where the notation implies that the first MC was used for construction of the $\Lambda$ operator (as described in \autoref{sec:theory}) and the remaining ten runs were production runs for calculation of the matrix elements. In all cases, the computational effort for the FD-GSIK was found to be weakly dependent on the system and exhibited sublinear scaling with respect to increasing system size. 
The standard deviation in the excitation energies were found to be two orders of magnitude smaller than the calculated averages. 
\begin{table}[]
\centering
\caption{The standard deviation of $\omega_{P}^{FD-GSIK}$ from 1$+$10 runs of the Monte Carlo integration with total single-core CPU computational time for the HOMO-LUMO particle-hole excitation.}
\label{tab:homo_lumo_Ag_std}
\begin{tabular}{cccccccc}
\textbf{Chemical} && $\omega_{P}^{FD-GSIK}$ && Standard && Total CPU  \\
\textbf{ Formula} && (eV) && Deviation (eV)  && Time (Hours)\\ \hline
Ag$_{2}$    & &	1.5835    & & 3.68E-03      & &	1.31                      \\
Ag$_{4}$     & &	1.2387    & & 1.57E-03      & &	1.35                      \\
Ag$_{6}$     & &	1.0766    & & 1.97E-03      & &	1.36                       \\
Ag$_{8}$     & &	0.8822    & & 3.52E-03      & &	1.35                       \\
Ag$_{10}$   & &	0.8690   & & 2.62E-03     & &	1.32                       \\
Ag$_{12}$   & &	0.7757    & & 4.75E-03     & &	1.33                       \\
Ag$_{20}$   & &	0.5217    & & 3.70E-03     & &	1.34                       \\
Ag$_{40}$   & &	0.4805   & & 6.96E-03     & &	1.32                        \\
Ag$_{60}$   & &	0.6695   & & 5.50E-03     & &	1.33                        \\
Ag$_{80}$   & &	0.8671   & & 4.44E-03     & &	1.35                        \\
Ag$_{100}$ & &	1.8456   & & 2.59E-03     & &	1.33                      \\ 
\end{tabular}
\end{table}
\begin{table}[]
\centering
\caption{The standard deviation of $\omega_{P}^{FD-GSIK}$ from 1$+$10 runs of the Monte Carlo integration with total single-core CPU computational time for the excitation in the dense manifold of particle-hole states. }
\label{tab:dressed_Ag_std}
\begin{tabular}{cccccccc}
\textbf{Chemical} && $\omega_{P}^{FD-GSIK}$ && Standard && Total CPU  \\
\textbf{ Formula} && (eV) && Deviation (eV)  && Time (Hours)\\ \hline
Ag$_{10}$ & &	    2.7217    & & 5.32E-04    & &	1.34   \\
Ag$_{20}$ & &	    2.5120    & & 9.68E-04    & &	1.36     \\
Ag$_{40}$  & &	2.3287    & & 3.38E-03    & &	1.34     \\
Ag$_{60}$  & &	2.7574    & & 1.30E-03    & &	1.35    \\
Ag$_{80}$  & &	2.8929    & & 1.56E-03    & &	1.34     \\
Ag$_{100}$ & &	2.6626    & & 1.84E-03    & &	1.34     \\
\end{tabular}
\end{table}

\section{Conclusions}\label{sec:conclusions}
In this work, a frequency-dependent explicitly correlated electron-hole interaction kernel (FD-GSIK)
method was developed for treating electronic excitation in a dense manifold of particle-hole states. 
The method is based on first constructing an optimized energy-restricted subspace of particle and hole states and then using an explicitly correlated ansatz for the correlated electron-hole wave function. The FD-GSIK method is specifically designed to avoid the steep computational cost associated with conventional approaches to excited state calculations of large nanoparticles. Specifically, FD-GSIK avoids the AO-to-MO two-electron integral transformations and calculates all the necessary MO integrals directly in real-space representation using Monte Carlo integration. The use of a large number of particle-hole states common in CIS and linear-response TDDFT calculation is also avoided by use of a frequency-dependent and explicitly correlated electron-hole interaction kernel. The developed method was benchmarked against EOM-CCSD and real-time TDDFT methods and was applied to investigations of the electronic excitation in a series of silver nanowires ($\mathrm{Ag}_n, n=2,\dots,100$). It was found that the low-energy electronic excitations in the nanowires followed nonmonotonic behavior with respect to increasing wire length and exhibited an inflection point at $n=50$. Future study of these nanowires will be based on the periodic-boundary implementation of the FD-GSIK method to investigative excitation in the infinitely long limit.
\par
\begin{acknowledgments}
This research was supported by the National Science Foundation under Grant No. CHE-1349892,
ACI-1341006, ACI-1541396 and by computational resources provided by Syracuse University. 
\end{acknowledgments}
\par
\textbf{Data Availability Statements:} The data that supports the findings of this study are available within the article (and its supplementary material).  Additional data is available from the corresponding author upon request.

\section{Appendix}\label{sec:appendix}
\begin{table}[h!]
\centering
\caption{The energy cut-off parameters for the collection of single-particle excitations for Ag nanowires with the number of particle/hole states in $P$ and $Q$.}
\label{tab:dressed_Ag_details}
\begin{tabular}{ccccccccccccc}
\textbf{Chemical  }     &   & $\eta_h$ &   & $N_{P^h}$ && $N_{Q^h}$ & & $\eta_e$   & & $N_{P^e}$ & &$N_{Q^e}$ \\ 
\textbf{Formula} & & (a.u.) && && && (a.u.) && && \\ \hline
Ag$_{10}$  && -2.12E-01 && 3    && 92    && -4.65E-02 & & 2   &  & 143   \\
Ag$_{20}$  && -2.11E-01 && 5     && 185  & & -4.31E-02 && 4   &  & 286   \\
Ag$_{40}$  && -2.11E-01 && 9    & & 371   && -4.86E-02 && 7    & & 573   \\
Ag$_{60}$  && -2.11E-01 && 13   & & 557   && -5.07E-02 && 10    && 860   \\
Ag$_{80}$ & & -2.11E-01 && 17   & & 743  & & -5.17E-02 && 13   & & 1147  \\
Ag$_{100}$ && -2.03E-01 && 17    && 933  & & -5.84E-02 && 14    && 1436  
\end{tabular}
\end{table}

\newpage
\bibliography{mybib}

\begin{thebibliography}{88}%
\makeatletter
\providecommand \@ifxundefined [1]{%
 \@ifx{#1\undefined}
}%
\providecommand \@ifnum [1]{%
 \ifnum #1\expandafter \@firstoftwo
 \else \expandafter \@secondoftwo
 \fi
}%
\providecommand \@ifx [1]{%
 \ifx #1\expandafter \@firstoftwo
 \else \expandafter \@secondoftwo
 \fi
}%
\providecommand \natexlab [1]{#1}%
\providecommand \enquote  [1]{``#1''}%
\providecommand \bibnamefont  [1]{#1}%
\providecommand \bibfnamefont [1]{#1}%
\providecommand \citenamefont [1]{#1}%
\providecommand \href@noop [0]{\@secondoftwo}%
\providecommand \href [0]{\begingroup \@sanitize@url \@href}%
\providecommand \@href[1]{\@@startlink{#1}\@@href}%
\providecommand \@@href[1]{\endgroup#1\@@endlink}%
\providecommand \@sanitize@url [0]{\catcode `\\12\catcode `\$12\catcode
  `\&12\catcode `\#12\catcode `\^12\catcode `\_12\catcode `\%12\relax}%
\providecommand \@@startlink[1]{}%
\providecommand \@@endlink[0]{}%
\providecommand \url  [0]{\begingroup\@sanitize@url \@url }%
\providecommand \@url [1]{\endgroup\@href {#1}{\urlprefix }}%
\providecommand \urlprefix  [0]{URL }%
\providecommand \Eprint [0]{\href }%
\providecommand \doibase [0]{http://dx.doi.org/}%
\providecommand \selectlanguage [0]{\@gobble}%
\providecommand \bibinfo  [0]{\@secondoftwo}%
\providecommand \bibfield  [0]{\@secondoftwo}%
\providecommand \translation [1]{[#1]}%
\providecommand \BibitemOpen [0]{}%
\providecommand \bibitemStop [0]{}%
\providecommand \bibitemNoStop [0]{.\EOS\space}%
\providecommand \EOS [0]{\spacefactor3000\relax}%
\providecommand \BibitemShut  [1]{\csname bibitem#1\endcsname}%
\let\auto@bib@innerbib\@empty
\bibitem [{\citenamefont {Haes}\ \emph {et~al.}(2004)\citenamefont {Haes},
  \citenamefont {Zou}, \citenamefont {Schatz},\ and\ \citenamefont
  {Van~Duyne}}]{Haes20046961}%
  \BibitemOpen
  \bibfield  {author} {\bibinfo {author} {\bibfnamefont {A.}~\bibnamefont
  {Haes}}, \bibinfo {author} {\bibfnamefont {S.}~\bibnamefont {Zou}}, \bibinfo
  {author} {\bibfnamefont {G.}~\bibnamefont {Schatz}}, \ and\ \bibinfo {author}
  {\bibfnamefont {R.}~\bibnamefont {Van~Duyne}},\ }\bibfield  {title} {\enquote
  {\bibinfo {title} {Nanoscale optical biosensor: Short range distance
  dependence of the localized surface plasmon resonance of noble metal
  nanoparticles},}\ }\href {\doibase 10.1021/jp036261n} {\bibfield  {journal}
  {\bibinfo  {journal} {Journal of Physical Chemistry B}\ }\textbf {\bibinfo
  {volume} {108}},\ \bibinfo {pages} {6961--6968} (\bibinfo {year}
  {2004})}\BibitemShut {NoStop}%
\bibitem [{\citenamefont {Govorov}\ and\ \citenamefont
  {Richardson}(2007)}]{Govorov200730}%
  \BibitemOpen
  \bibfield  {author} {\bibinfo {author} {\bibfnamefont {A.}~\bibnamefont
  {Govorov}}\ and\ \bibinfo {author} {\bibfnamefont {H.}~\bibnamefont
  {Richardson}},\ }\bibfield  {title} {\enquote {\bibinfo {title} {Generating
  heat with metal nanoparticles},}\ }\href {\doibase
  10.1016/S1748-0132(07)70017-8} {\bibfield  {journal} {\bibinfo  {journal}
  {Nano Today}\ }\textbf {\bibinfo {volume} {2}},\ \bibinfo {pages} {30--38}
  (\bibinfo {year} {2007})}\BibitemShut {NoStop}%
\bibitem [{\citenamefont {Tallury}\ \emph {et~al.}(2010)\citenamefont
  {Tallury}, \citenamefont {Malhotra}, \citenamefont {Byrne},\ and\
  \citenamefont {Santra}}]{Tallury2010424}%
  \BibitemOpen
  \bibfield  {author} {\bibinfo {author} {\bibfnamefont {P.}~\bibnamefont
  {Tallury}}, \bibinfo {author} {\bibfnamefont {A.}~\bibnamefont {Malhotra}},
  \bibinfo {author} {\bibfnamefont {L.}~\bibnamefont {Byrne}}, \ and\ \bibinfo
  {author} {\bibfnamefont {S.}~\bibnamefont {Santra}},\ }\bibfield  {title}
  {\enquote {\bibinfo {title} {Nanobioimaging and sensing of infectious
  diseases},}\ }\href {\doibase 10.1016/j.addr.2009.11.014} {\bibfield
  {journal} {\bibinfo  {journal} {Advanced Drug Delivery Reviews}\ }\textbf
  {\bibinfo {volume} {62}},\ \bibinfo {pages} {424--437} (\bibinfo {year}
  {2010})}\BibitemShut {NoStop}%
\bibitem [{\citenamefont {Sukirtha}\ \emph {et~al.}(2012)\citenamefont
  {Sukirtha}, \citenamefont {Priyanka}, \citenamefont {Antony}, \citenamefont
  {Kamalakkannan}, \citenamefont {Thangam}, \citenamefont {Gunasekaran},
  \citenamefont {Krishnan},\ and\ \citenamefont {Achiraman}}]{Sukirtha2012273}%
  \BibitemOpen
  \bibfield  {author} {\bibinfo {author} {\bibfnamefont {R.}~\bibnamefont
  {Sukirtha}}, \bibinfo {author} {\bibfnamefont {K.}~\bibnamefont {Priyanka}},
  \bibinfo {author} {\bibfnamefont {J.}~\bibnamefont {Antony}}, \bibinfo
  {author} {\bibfnamefont {S.}~\bibnamefont {Kamalakkannan}}, \bibinfo {author}
  {\bibfnamefont {R.}~\bibnamefont {Thangam}}, \bibinfo {author} {\bibfnamefont
  {P.}~\bibnamefont {Gunasekaran}}, \bibinfo {author} {\bibfnamefont
  {M.}~\bibnamefont {Krishnan}}, \ and\ \bibinfo {author} {\bibfnamefont
  {S.}~\bibnamefont {Achiraman}},\ }\bibfield  {title} {\enquote {\bibinfo
  {title} {Cytotoxic effect of green synthesized silver nanoparticles using
  melia azedarach against in vitro hela cell lines and lymphoma mice model},}\
  }\href {\doibase 10.1016/j.procbio.2011.11.003} {\bibfield  {journal}
  {\bibinfo  {journal} {Process Biochemistry}\ }\textbf {\bibinfo {volume}
  {47}},\ \bibinfo {pages} {273--279} (\bibinfo {year} {2012})}\BibitemShut
  {NoStop}%
\bibitem [{\citenamefont {Khlebtsov}\ and\ \citenamefont
  {Dykman}(2010)}]{Khlebtsov20101}%
  \BibitemOpen
  \bibfield  {author} {\bibinfo {author} {\bibfnamefont {N.}~\bibnamefont
  {Khlebtsov}}\ and\ \bibinfo {author} {\bibfnamefont {L.}~\bibnamefont
  {Dykman}},\ }\bibfield  {title} {\enquote {\bibinfo {title} {Optical
  properties and biomedical applications of plasmonic nanoparticles},}\ }\href
  {\doibase 10.1016/j.jqsrt.2009.07.012} {\bibfield  {journal} {\bibinfo
  {journal} {Journal of Quantitative Spectroscopy and Radiative Transfer}\
  }\textbf {\bibinfo {volume} {111}},\ \bibinfo {pages} {1--35} (\bibinfo
  {year} {2010})}\BibitemShut {NoStop}%
\bibitem [{\citenamefont {Kumar}\ \emph {et~al.}(2013)\citenamefont {Kumar},
  \citenamefont {Govindaraju}, \citenamefont {Senthamilselvi},\ and\
  \citenamefont {Premkumar}}]{Kumar2013658}%
  \BibitemOpen
  \bibfield  {author} {\bibinfo {author} {\bibfnamefont {P.}~\bibnamefont
  {Kumar}}, \bibinfo {author} {\bibfnamefont {M.}~\bibnamefont {Govindaraju}},
  \bibinfo {author} {\bibfnamefont {S.}~\bibnamefont {Senthamilselvi}}, \ and\
  \bibinfo {author} {\bibfnamefont {K.}~\bibnamefont {Premkumar}},\ }\bibfield
  {title} {\enquote {\bibinfo {title} {Photocatalytic degradation of methyl
  orange dye using silver (ag) nanoparticles synthesized from ulva lactuca},}\
  }\href {\doibase 10.1016/j.colsurfb.2012.11.022} {\bibfield  {journal}
  {\bibinfo  {journal} {Colloids and Surfaces B: Biointerfaces}\ }\textbf
  {\bibinfo {volume} {103}},\ \bibinfo {pages} {658--661} (\bibinfo {year}
  {2013})}\BibitemShut {NoStop}%
\bibitem [{\citenamefont {Beck}, \citenamefont {Polman},\ and\ \citenamefont
  {Catchpole}(2009)}]{Beck2009}%
  \BibitemOpen
  \bibfield  {author} {\bibinfo {author} {\bibfnamefont {F.}~\bibnamefont
  {Beck}}, \bibinfo {author} {\bibfnamefont {A.}~\bibnamefont {Polman}}, \ and\
  \bibinfo {author} {\bibfnamefont {K.}~\bibnamefont {Catchpole}},\ }\bibfield
  {title} {\enquote {\bibinfo {title} {Tunable light trapping for solar cells
  using localized surface plasmons},}\ }\href {\doibase 10.1063/1.3140609}
  {\bibfield  {journal} {\bibinfo  {journal} {Journal of Applied Physics}\
  }\textbf {\bibinfo {volume} {105}} (\bibinfo {year} {2009}),\
  10.1063/1.3140609}\BibitemShut {NoStop}%
\bibitem [{\citenamefont {Kang}\ \emph {et~al.}(2010)\citenamefont {Kang},
  \citenamefont {Xu}, \citenamefont {Park}, \citenamefont {Luo},\ and\
  \citenamefont {Guo}}]{Kang20104378}%
  \BibitemOpen
  \bibfield  {author} {\bibinfo {author} {\bibfnamefont {M.-G.}\ \bibnamefont
  {Kang}}, \bibinfo {author} {\bibfnamefont {T.}~\bibnamefont {Xu}}, \bibinfo
  {author} {\bibfnamefont {H.}~\bibnamefont {Park}}, \bibinfo {author}
  {\bibfnamefont {X.}~\bibnamefont {Luo}}, \ and\ \bibinfo {author}
  {\bibfnamefont {L.}~\bibnamefont {Guo}},\ }\bibfield  {title} {\enquote
  {\bibinfo {title} {Efficiency enhancement of organic solar cells using
  transparent plasmonic ag nanowire electrodes},}\ }\href {\doibase
  10.1002/adma.201001395} {\bibfield  {journal} {\bibinfo  {journal} {Advanced
  Materials}\ }\textbf {\bibinfo {volume} {22}},\ \bibinfo {pages} {4378--4383}
  (\bibinfo {year} {2010})}\BibitemShut {NoStop}%
\bibitem [{\citenamefont {Xiong}\ \emph {et~al.}(2012)\citenamefont {Xiong},
  \citenamefont {Long}, \citenamefont {Liu}, \citenamefont {Zhong},
  \citenamefont {Wang}, \citenamefont {Li},\ and\ \citenamefont
  {Xie}}]{Xiong20124416}%
  \BibitemOpen
  \bibfield  {author} {\bibinfo {author} {\bibfnamefont {Y.}~\bibnamefont
  {Xiong}}, \bibinfo {author} {\bibfnamefont {R.}~\bibnamefont {Long}},
  \bibinfo {author} {\bibfnamefont {D.}~\bibnamefont {Liu}}, \bibinfo {author}
  {\bibfnamefont {X.}~\bibnamefont {Zhong}}, \bibinfo {author} {\bibfnamefont
  {C.}~\bibnamefont {Wang}}, \bibinfo {author} {\bibfnamefont {Z.-Y.}\
  \bibnamefont {Li}}, \ and\ \bibinfo {author} {\bibfnamefont {Y.}~\bibnamefont
  {Xie}},\ }\bibfield  {title} {\enquote {\bibinfo {title} {Solar energy
  conversion with tunable plasmonic nanostructures for thermoelectric
  devices},}\ }\href {\doibase 10.1039/c2nr30208j} {\bibfield  {journal}
  {\bibinfo  {journal} {Nanoscale}\ }\textbf {\bibinfo {volume} {4}},\ \bibinfo
  {pages} {4416--4420} (\bibinfo {year} {2012})}\BibitemShut {NoStop}%
\bibitem [{\citenamefont {Zengin}\ \emph {et~al.}(2015)\citenamefont {Zengin},
  \citenamefont {Wersäll}, \citenamefont {Nilsson}, \citenamefont
  {Antosiewicz}, \citenamefont {Käll},\ and\ \citenamefont
  {Shegai}}]{Zengin2015}%
  \BibitemOpen
  \bibfield  {author} {\bibinfo {author} {\bibfnamefont {G.}~\bibnamefont
  {Zengin}}, \bibinfo {author} {\bibfnamefont {M.}~\bibnamefont {Wersäll}},
  \bibinfo {author} {\bibfnamefont {S.}~\bibnamefont {Nilsson}}, \bibinfo
  {author} {\bibfnamefont {T.}~\bibnamefont {Antosiewicz}}, \bibinfo {author}
  {\bibfnamefont {M.}~\bibnamefont {Käll}}, \ and\ \bibinfo {author}
  {\bibfnamefont {T.}~\bibnamefont {Shegai}},\ }\bibfield  {title} {\enquote
  {\bibinfo {title} {Realizing strong light-matter interactions between
  single-nanoparticle plasmons and molecular excitons at ambient conditions},}\
  }\href {\doibase 10.1103/PhysRevLett.114.157401} {\bibfield  {journal}
  {\bibinfo  {journal} {Physical Review Letters}\ }\textbf {\bibinfo {volume}
  {114}} (\bibinfo {year} {2015}),\ 10.1103/PhysRevLett.114.157401}\BibitemShut
  {NoStop}%
\bibitem [{\citenamefont {West}\ and\ \citenamefont
  {Halas}(2003)}]{West2003285}%
  \BibitemOpen
  \bibfield  {author} {\bibinfo {author} {\bibfnamefont {J.}~\bibnamefont
  {West}}\ and\ \bibinfo {author} {\bibfnamefont {N.}~\bibnamefont {Halas}},\
  }\bibfield  {title} {\enquote {\bibinfo {title} {Engineered nanomaterials for
  biophotonics applications: Improving sensing, imaging, and therapeutics},}\
  }\href {\doibase 10.1146/annurev.bioeng.5.011303.120723} {\bibfield
  {journal} {\bibinfo  {journal} {Annual Review of Biomedical Engineering}\
  }\textbf {\bibinfo {volume} {5}},\ \bibinfo {pages} {285--292} (\bibinfo
  {year} {2003})}\BibitemShut {NoStop}%
\bibitem [{\citenamefont {McFarland}\ and\ \citenamefont
  {Van~Duyne}(2003)}]{McFarland20031057}%
  \BibitemOpen
  \bibfield  {author} {\bibinfo {author} {\bibfnamefont {A.}~\bibnamefont
  {McFarland}}\ and\ \bibinfo {author} {\bibfnamefont {R.}~\bibnamefont
  {Van~Duyne}},\ }\bibfield  {title} {\enquote {\bibinfo {title} {Single silver
  nanoparticles as real-time optical sensors with zeptomole sensitivity},}\
  }\href {\doibase 10.1021/nl034372s} {\bibfield  {journal} {\bibinfo
  {journal} {Nano Letters}\ }\textbf {\bibinfo {volume} {3}},\ \bibinfo {pages}
  {1057--1062} (\bibinfo {year} {2003})}\BibitemShut {NoStop}%
\bibitem [{\citenamefont {Jain}\ \emph {et~al.}(2008)\citenamefont {Jain},
  \citenamefont {Huang}, \citenamefont {El-Sayed},\ and\ \citenamefont
  {El-Sayed}}]{Jain20081578}%
  \BibitemOpen
  \bibfield  {author} {\bibinfo {author} {\bibfnamefont {P.}~\bibnamefont
  {Jain}}, \bibinfo {author} {\bibfnamefont {X.}~\bibnamefont {Huang}},
  \bibinfo {author} {\bibfnamefont {I.}~\bibnamefont {El-Sayed}}, \ and\
  \bibinfo {author} {\bibfnamefont {M.}~\bibnamefont {El-Sayed}},\ }\bibfield
  {title} {\enquote {\bibinfo {title} {Noble metals on the nanoscale: Optical
  and photothermal properties and some applications in imaging, sensing,
  biology, and medicine},}\ }\href {\doibase 10.1021/ar7002804} {\bibfield
  {journal} {\bibinfo  {journal} {Accounts of Chemical Research}\ }\textbf
  {\bibinfo {volume} {41}},\ \bibinfo {pages} {1578--1586} (\bibinfo {year}
  {2008})}\BibitemShut {NoStop}%
\bibitem [{\citenamefont {Feng}\ \emph {et~al.}(2012)\citenamefont {Feng},
  \citenamefont {Siu}, \citenamefont {Roelke}, \citenamefont {Mehta},
  \citenamefont {Rhieu}, \citenamefont {Palmore},\ and\ \citenamefont
  {Pacifici}}]{Feng2012602}%
  \BibitemOpen
  \bibfield  {author} {\bibinfo {author} {\bibfnamefont {J.}~\bibnamefont
  {Feng}}, \bibinfo {author} {\bibfnamefont {V.}~\bibnamefont {Siu}}, \bibinfo
  {author} {\bibfnamefont {A.}~\bibnamefont {Roelke}}, \bibinfo {author}
  {\bibfnamefont {V.}~\bibnamefont {Mehta}}, \bibinfo {author} {\bibfnamefont
  {S.}~\bibnamefont {Rhieu}}, \bibinfo {author} {\bibfnamefont
  {G.}~\bibnamefont {Palmore}}, \ and\ \bibinfo {author} {\bibfnamefont
  {D.}~\bibnamefont {Pacifici}},\ }\bibfield  {title} {\enquote {\bibinfo
  {title} {Nanoscale plasmonic interferometers for multispectral,
  high-throughput biochemical sensing},}\ }\href {\doibase 10.1021/nl203325s}
  {\bibfield  {journal} {\bibinfo  {journal} {Nano Letters}\ }\textbf {\bibinfo
  {volume} {12}},\ \bibinfo {pages} {602--609} (\bibinfo {year}
  {2012})}\BibitemShut {NoStop}%
\bibitem [{\citenamefont {L\'{o}pez~Lozano}, \citenamefont {Mottet},\ and\
  \citenamefont {Weissker}(2013)}]{LopezLozano20133062}%
  \BibitemOpen
  \bibfield  {author} {\bibinfo {author} {\bibfnamefont {X.}~\bibnamefont
  {L\'{o}pez~Lozano}}, \bibinfo {author} {\bibfnamefont {C.}~\bibnamefont
  {Mottet}}, \ and\ \bibinfo {author} {\bibfnamefont {H.-C.}\ \bibnamefont
  {Weissker}},\ }\bibfield  {title} {\enquote {\bibinfo {title} {Effect of
  alloying on the optical properties of ag-au nanoparticles},}\ }\href
  {\doibase 10.1021/jp309957y} {\bibfield  {journal} {\bibinfo  {journal}
  {Journal of Physical Chemistry C}\ }\textbf {\bibinfo {volume} {117}},\
  \bibinfo {pages} {3062--3068} (\bibinfo {year} {2013})}\BibitemShut {NoStop}%
\bibitem [{\citenamefont {Marinica}\ \emph {et~al.}(2012)\citenamefont
  {Marinica}, \citenamefont {Kazansky}, \citenamefont {Nordlander},
  \citenamefont {Aizpurua},\ and\ \citenamefont {Borisov}}]{Marinica20121333}%
  \BibitemOpen
  \bibfield  {author} {\bibinfo {author} {\bibfnamefont {D.}~\bibnamefont
  {Marinica}}, \bibinfo {author} {\bibfnamefont {A.}~\bibnamefont {Kazansky}},
  \bibinfo {author} {\bibfnamefont {P.}~\bibnamefont {Nordlander}}, \bibinfo
  {author} {\bibfnamefont {J.}~\bibnamefont {Aizpurua}}, \ and\ \bibinfo
  {author} {\bibfnamefont {A.}~\bibnamefont {Borisov}},\ }\bibfield  {title}
  {\enquote {\bibinfo {title} {Quantum plasmonics: Nonlinear effects in the
  field enhancement of a plasmonic nanoparticle dimer},}\ }\href {\doibase
  10.1021/nl300269c} {\bibfield  {journal} {\bibinfo  {journal} {Nano Letters}\
  }\textbf {\bibinfo {volume} {12}},\ \bibinfo {pages} {1333--1339} (\bibinfo
  {year} {2012})}\BibitemShut {NoStop}%
\bibitem [{\citenamefont {Malinsky}\ \emph {et~al.}(2001)\citenamefont
  {Malinsky}, \citenamefont {Kelly}, \citenamefont {Schatz},\ and\
  \citenamefont {Van~Duyne}}]{Malinsky20011471}%
  \BibitemOpen
  \bibfield  {author} {\bibinfo {author} {\bibfnamefont {M.}~\bibnamefont
  {Malinsky}}, \bibinfo {author} {\bibfnamefont {K.}~\bibnamefont {Kelly}},
  \bibinfo {author} {\bibfnamefont {G.}~\bibnamefont {Schatz}}, \ and\ \bibinfo
  {author} {\bibfnamefont {R.}~\bibnamefont {Van~Duyne}},\ }\bibfield  {title}
  {\enquote {\bibinfo {title} {Chain length dependence and sensing capabilities
  of the localized surface plasmon resonance of silver nanoparticles chemically
  modified with alkanethiol self-assembled monolayers},}\ }\href {\doibase
  10.1021/ja003312a} {\bibfield  {journal} {\bibinfo  {journal} {Journal of the
  American Chemical Society}\ }\textbf {\bibinfo {volume} {123}},\ \bibinfo
  {pages} {1471--1482} (\bibinfo {year} {2001})}\BibitemShut {NoStop}%
\bibitem [{\citenamefont {Aikens}, \citenamefont {Li},\ and\ \citenamefont
  {Schatz}(2008)}]{Aikens200811272}%
  \BibitemOpen
  \bibfield  {author} {\bibinfo {author} {\bibfnamefont {C.}~\bibnamefont
  {Aikens}}, \bibinfo {author} {\bibfnamefont {S.}~\bibnamefont {Li}}, \ and\
  \bibinfo {author} {\bibfnamefont {G.}~\bibnamefont {Schatz}},\ }\bibfield
  {title} {\enquote {\bibinfo {title} {From discrete electronic states to
  plasmons: Tddft optical absorption properties of agn (n = 10, 20, 35, 56, 84,
  120) tetrahedral clusters},}\ }\href {\doibase 10.1021/jp802707r} {\bibfield
  {journal} {\bibinfo  {journal} {Journal of Physical Chemistry C}\ }\textbf
  {\bibinfo {volume} {112}},\ \bibinfo {pages} {11272--11279} (\bibinfo {year}
  {2008})}\BibitemShut {NoStop}%
\bibitem [{\citenamefont {Kelly}\ \emph {et~al.}(2003)\citenamefont {Kelly},
  \citenamefont {Coronado}, \citenamefont {Zhao},\ and\ \citenamefont
  {Schatz}}]{Kelly2003668}%
  \BibitemOpen
  \bibfield  {author} {\bibinfo {author} {\bibfnamefont {K.}~\bibnamefont
  {Kelly}}, \bibinfo {author} {\bibfnamefont {E.}~\bibnamefont {Coronado}},
  \bibinfo {author} {\bibfnamefont {L.}~\bibnamefont {Zhao}}, \ and\ \bibinfo
  {author} {\bibfnamefont {G.}~\bibnamefont {Schatz}},\ }\bibfield  {title}
  {\enquote {\bibinfo {title} {The optical properties of metal nanoparticles:
  The influence of size, shape, and dielectric environment},}\ }\href {\doibase
  10.1021/jp026731y} {\bibfield  {journal} {\bibinfo  {journal} {Journal of
  Physical Chemistry B}\ }\textbf {\bibinfo {volume} {107}},\ \bibinfo {pages}
  {668--677} (\bibinfo {year} {2003})}\BibitemShut {NoStop}%
\bibitem [{\citenamefont {Ross}, \citenamefont {Mirkin},\ and\ \citenamefont
  {Schatz}(2016)}]{Ross2016816}%
  \BibitemOpen
  \bibfield  {author} {\bibinfo {author} {\bibfnamefont {M.}~\bibnamefont
  {Ross}}, \bibinfo {author} {\bibfnamefont {C.}~\bibnamefont {Mirkin}}, \ and\
  \bibinfo {author} {\bibfnamefont {G.}~\bibnamefont {Schatz}},\ }\bibfield
  {title} {\enquote {\bibinfo {title} {Optical properties of one-, two-, and
  three-dimensional arrays of plasmonic nanostructures},}\ }\href {\doibase
  10.1021/acs.jpcc.5b10800} {\bibfield  {journal} {\bibinfo  {journal} {Journal
  of Physical Chemistry C}\ }\textbf {\bibinfo {volume} {120}},\ \bibinfo
  {pages} {816--830} (\bibinfo {year} {2016})}\BibitemShut {NoStop}%
\bibitem [{\citenamefont {Shabaninezhad}\ and\ \citenamefont
  {Ramakrishna}(2019)}]{Shabaninezhad2019}%
  \BibitemOpen
  \bibfield  {author} {\bibinfo {author} {\bibfnamefont {M.}~\bibnamefont
  {Shabaninezhad}}\ and\ \bibinfo {author} {\bibfnamefont {G.}~\bibnamefont
  {Ramakrishna}},\ }\bibfield  {title} {\enquote {\bibinfo {title} {Theoretical
  investigation of size, shape, and aspect ratio effect on the lspr sensitivity
  of hollow-gold nanoshells},}\ }\href {\doibase 10.1063/1.5090885} {\bibfield
  {journal} {\bibinfo  {journal} {Journal of Chemical Physics}\ }\textbf
  {\bibinfo {volume} {150}} (\bibinfo {year} {2019}),\
  10.1063/1.5090885}\BibitemShut {NoStop}%
\bibitem [{\citenamefont {Bae}\ and\ \citenamefont
  {Aikens}(2012{\natexlab{a}})}]{Bae201210356}%
  \BibitemOpen
  \bibfield  {author} {\bibinfo {author} {\bibfnamefont {G.-T.}\ \bibnamefont
  {Bae}}\ and\ \bibinfo {author} {\bibfnamefont {C.}~\bibnamefont {Aikens}},\
  }\bibfield  {title} {\enquote {\bibinfo {title} {Time-dependent density
  functional theory studies of optical properties of ag nanoparticles:
  Octahedra, truncated octahedra, and icosahedra},}\ }\href {\doibase
  10.1021/jp300789x} {\bibfield  {journal} {\bibinfo  {journal} {Journal of
  Physical Chemistry C}\ }\textbf {\bibinfo {volume} {116}},\ \bibinfo {pages}
  {10356--10367} (\bibinfo {year} {2012}{\natexlab{a}})}\BibitemShut {NoStop}%
\bibitem [{\citenamefont {Eustis}\ and\ \citenamefont
  {El-Sayed}(2006)}]{Eustis2006209}%
  \BibitemOpen
  \bibfield  {author} {\bibinfo {author} {\bibfnamefont {S.}~\bibnamefont
  {Eustis}}\ and\ \bibinfo {author} {\bibfnamefont {M.}~\bibnamefont
  {El-Sayed}},\ }\bibfield  {title} {\enquote {\bibinfo {title} {Why gold
  nanoparticles are more precious than pretty gold: Noble metal surface plasmon
  resonance and its enhancement of the radiative and nonradiative properties of
  nanocrystals of different shapes},}\ }\href {\doibase 10.1039/b514191e}
  {\bibfield  {journal} {\bibinfo  {journal} {Chemical Society Reviews}\
  }\textbf {\bibinfo {volume} {35}},\ \bibinfo {pages} {209--217} (\bibinfo
  {year} {2006})}\BibitemShut {NoStop}%
\bibitem [{\citenamefont {Liz-Marz\'{a}n}(2006)}]{LizMarzan200632}%
  \BibitemOpen
  \bibfield  {author} {\bibinfo {author} {\bibfnamefont {L.}~\bibnamefont
  {Liz-Marz\'{a}n}},\ }\bibfield  {title} {\enquote {\bibinfo {title}
  {Tailoring surface plasmons through the morphology and assembly of metal
  nanoparticles},}\ }\href {\doibase 10.1021/la0513353} {\bibfield  {journal}
  {\bibinfo  {journal} {Langmuir}\ }\textbf {\bibinfo {volume} {22}},\ \bibinfo
  {pages} {32--41} (\bibinfo {year} {2006})}\BibitemShut {NoStop}%
\bibitem [{\citenamefont {Sonnefraud}\ \emph {et~al.}(2012)\citenamefont
  {Sonnefraud}, \citenamefont {Leen~Koh}, \citenamefont {McComb},\ and\
  \citenamefont {Maier}}]{Sonnefraud2012277}%
  \BibitemOpen
  \bibfield  {author} {\bibinfo {author} {\bibfnamefont {Y.}~\bibnamefont
  {Sonnefraud}}, \bibinfo {author} {\bibfnamefont {A.}~\bibnamefont
  {Leen~Koh}}, \bibinfo {author} {\bibfnamefont {D.}~\bibnamefont {McComb}}, \
  and\ \bibinfo {author} {\bibfnamefont {S.}~\bibnamefont {Maier}},\ }\bibfield
   {title} {\enquote {\bibinfo {title} {Nanoplasmonics: Engineering and
  observation of localized plasmon modes},}\ }\href {\doibase
  10.1002/lpor.201100027} {\bibfield  {journal} {\bibinfo  {journal} {Laser and
  Photonics Reviews}\ }\textbf {\bibinfo {volume} {6}},\ \bibinfo {pages}
  {277--295} (\bibinfo {year} {2012})}\BibitemShut {NoStop}%
\bibitem [{\citenamefont {Halas}\ \emph {et~al.}(2011)\citenamefont {Halas},
  \citenamefont {Lal}, \citenamefont {Chang}, \citenamefont {Link},\ and\
  \citenamefont {Nordlander}}]{Halas20113913}%
  \BibitemOpen
  \bibfield  {author} {\bibinfo {author} {\bibfnamefont {N.}~\bibnamefont
  {Halas}}, \bibinfo {author} {\bibfnamefont {S.}~\bibnamefont {Lal}}, \bibinfo
  {author} {\bibfnamefont {W.-S.}\ \bibnamefont {Chang}}, \bibinfo {author}
  {\bibfnamefont {S.}~\bibnamefont {Link}}, \ and\ \bibinfo {author}
  {\bibfnamefont {P.}~\bibnamefont {Nordlander}},\ }\bibfield  {title}
  {\enquote {\bibinfo {title} {Plasmons in strongly coupled metallic
  nanostructures},}\ }\href {\doibase 10.1021/cr200061k} {\bibfield  {journal}
  {\bibinfo  {journal} {Chemical Reviews}\ }\textbf {\bibinfo {volume} {111}},\
  \bibinfo {pages} {3913--3961} (\bibinfo {year} {2011})}\BibitemShut {NoStop}%
\bibitem [{\citenamefont {Geisler}\ \emph {et~al.}(2017)\citenamefont
  {Geisler}, \citenamefont {Krauss}, \citenamefont {Razinskas},\ and\
  \citenamefont {Hecht}}]{Geisler20171615}%
  \BibitemOpen
  \bibfield  {author} {\bibinfo {author} {\bibfnamefont {P.}~\bibnamefont
  {Geisler}}, \bibinfo {author} {\bibfnamefont {E.}~\bibnamefont {Krauss}},
  \bibinfo {author} {\bibfnamefont {G.}~\bibnamefont {Razinskas}}, \ and\
  \bibinfo {author} {\bibfnamefont {B.}~\bibnamefont {Hecht}},\ }\bibfield
  {title} {\enquote {\bibinfo {title} {Transmission of plasmons through a
  nanowire},}\ }\href {\doibase 10.1021/acsphotonics.7b00292} {\bibfield
  {journal} {\bibinfo  {journal} {ACS Photonics}\ }\textbf {\bibinfo {volume}
  {4}},\ \bibinfo {pages} {1615--1620} (\bibinfo {year} {2017})}\BibitemShut
  {NoStop}%
\bibitem [{\citenamefont {Knight}\ \emph {et~al.}(2007)\citenamefont {Knight},
  \citenamefont {Grady}, \citenamefont {Bardhan}, \citenamefont {Hao},
  \citenamefont {Nordlander},\ and\ \citenamefont {Halas}}]{Knight20072346}%
  \BibitemOpen
  \bibfield  {author} {\bibinfo {author} {\bibfnamefont {M.}~\bibnamefont
  {Knight}}, \bibinfo {author} {\bibfnamefont {N.}~\bibnamefont {Grady}},
  \bibinfo {author} {\bibfnamefont {R.}~\bibnamefont {Bardhan}}, \bibinfo
  {author} {\bibfnamefont {F.}~\bibnamefont {Hao}}, \bibinfo {author}
  {\bibfnamefont {P.}~\bibnamefont {Nordlander}}, \ and\ \bibinfo {author}
  {\bibfnamefont {N.}~\bibnamefont {Halas}},\ }\bibfield  {title} {\enquote
  {\bibinfo {title} {Nanoparticle-mediated coupling of light into a
  nanowire},}\ }\href {\doibase 10.1021/nl071001t} {\bibfield  {journal}
  {\bibinfo  {journal} {Nano Letters}\ }\textbf {\bibinfo {volume} {7}},\
  \bibinfo {pages} {2346--2350} (\bibinfo {year} {2007})}\BibitemShut {NoStop}%
\bibitem [{\citenamefont {Weerawardene}, \citenamefont {Häkkinen},\ and\
  \citenamefont {Aikens}(2018)}]{Weerawardene2018205}%
  \BibitemOpen
  \bibfield  {author} {\bibinfo {author} {\bibfnamefont {K.}~\bibnamefont
  {Weerawardene}}, \bibinfo {author} {\bibfnamefont {H.}~\bibnamefont
  {Häkkinen}}, \ and\ \bibinfo {author} {\bibfnamefont {C.}~\bibnamefont
  {Aikens}},\ }\bibfield  {title} {\enquote {\bibinfo {title} {Connections
  between theory and experiment for gold and silver nanoclusters},}\ }\href
  {\doibase 10.1146/annurev-physchem-052516-050932} {\bibfield  {journal}
  {\bibinfo  {journal} {Annual Review of Physical Chemistry}\ }\textbf
  {\bibinfo {volume} {69}},\ \bibinfo {pages} {205--229} (\bibinfo {year}
  {2018})}\BibitemShut {NoStop}%
\bibitem [{\citenamefont {Jain}\ and\ \citenamefont
  {El-Sayed}(2010)}]{Jain2010153}%
  \BibitemOpen
  \bibfield  {author} {\bibinfo {author} {\bibfnamefont {P.}~\bibnamefont
  {Jain}}\ and\ \bibinfo {author} {\bibfnamefont {M.}~\bibnamefont
  {El-Sayed}},\ }\bibfield  {title} {\enquote {\bibinfo {title} {Plasmonic
  coupling in noble metal nanostructures},}\ }\href {\doibase
  10.1016/j.cplett.2010.01.062} {\bibfield  {journal} {\bibinfo  {journal}
  {Chemical Physics Letters}\ }\textbf {\bibinfo {volume} {487}},\ \bibinfo
  {pages} {153--164} (\bibinfo {year} {2010})}\BibitemShut {NoStop}%
\bibitem [{\citenamefont {Varas}\ \emph {et~al.}(2016)\citenamefont {Varas},
  \citenamefont {García-González}, \citenamefont {Feist}, \citenamefont
  {García-Vidal},\ and\ \citenamefont {Rubio}}]{Varas2016409}%
  \BibitemOpen
  \bibfield  {author} {\bibinfo {author} {\bibfnamefont {A.}~\bibnamefont
  {Varas}}, \bibinfo {author} {\bibfnamefont {P.}~\bibnamefont
  {García-González}}, \bibinfo {author} {\bibfnamefont {J.}~\bibnamefont
  {Feist}}, \bibinfo {author} {\bibfnamefont {F.}~\bibnamefont
  {García-Vidal}}, \ and\ \bibinfo {author} {\bibfnamefont {A.}~\bibnamefont
  {Rubio}},\ }\bibfield  {title} {\enquote {\bibinfo {title} {Quantum
  plasmonics: from jellium models to ab initio calculations},}\ }\href
  {\doibase 10.1515/nanoph-2015-0141} {\bibfield  {journal} {\bibinfo
  {journal} {Nanophotonics}\ }\textbf {\bibinfo {volume} {5}},\ \bibinfo
  {pages} {409--426} (\bibinfo {year} {2016})}\BibitemShut {NoStop}%
\bibitem [{\citenamefont {Morton}, \citenamefont {Silverstein},\ and\
  \citenamefont {Jensen}(2011)}]{Morton20113962}%
  \BibitemOpen
  \bibfield  {author} {\bibinfo {author} {\bibfnamefont {S.}~\bibnamefont
  {Morton}}, \bibinfo {author} {\bibfnamefont {D.}~\bibnamefont {Silverstein}},
  \ and\ \bibinfo {author} {\bibfnamefont {L.}~\bibnamefont {Jensen}},\
  }\bibfield  {title} {\enquote {\bibinfo {title} {Theoretical studies of
  plasmonics using electronic structure methods},}\ }\href {\doibase
  10.1021/cr100265f} {\bibfield  {journal} {\bibinfo  {journal} {Chemical
  Reviews}\ }\textbf {\bibinfo {volume} {111}},\ \bibinfo {pages} {3962--3994}
  (\bibinfo {year} {2011})}\BibitemShut {NoStop}%
\bibitem [{\citenamefont {Bernadotte}, \citenamefont {Evers},\ and\
  \citenamefont {Jacob}(2013)}]{Bernadotte20131863}%
  \BibitemOpen
  \bibfield  {author} {\bibinfo {author} {\bibfnamefont {S.}~\bibnamefont
  {Bernadotte}}, \bibinfo {author} {\bibfnamefont {F.}~\bibnamefont {Evers}}, \
  and\ \bibinfo {author} {\bibfnamefont {C.}~\bibnamefont {Jacob}},\ }\bibfield
   {title} {\enquote {\bibinfo {title} {Plasmons in molecules},}\ }\href
  {\doibase 10.1021/jp3113073} {\bibfield  {journal} {\bibinfo  {journal}
  {Journal of Physical Chemistry C}\ }\textbf {\bibinfo {volume} {117}},\
  \bibinfo {pages} {1863--1878} (\bibinfo {year} {2013})}\BibitemShut {NoStop}%
\bibitem [{\citenamefont {Johnson}\ and\ \citenamefont
  {Aikens}(2009)}]{Johnson20094445}%
  \BibitemOpen
  \bibfield  {author} {\bibinfo {author} {\bibfnamefont {H.}~\bibnamefont
  {Johnson}}\ and\ \bibinfo {author} {\bibfnamefont {C.}~\bibnamefont
  {Aikens}},\ }\bibfield  {title} {\enquote {\bibinfo {title} {Electronic
  structure and tddft optical absorption spectra of silver nanorods},}\ }\href
  {\doibase 10.1021/jp811075u} {\bibfield  {journal} {\bibinfo  {journal}
  {Journal of Physical Chemistry A}\ }\textbf {\bibinfo {volume} {113}},\
  \bibinfo {pages} {4445--4450} (\bibinfo {year} {2009})}\BibitemShut {NoStop}%
\bibitem [{\citenamefont {Gao}, \citenamefont {Ruud},\ and\ \citenamefont
  {Luo}(2014)}]{Gao201413059}%
  \BibitemOpen
  \bibfield  {author} {\bibinfo {author} {\bibfnamefont {B.}~\bibnamefont
  {Gao}}, \bibinfo {author} {\bibfnamefont {K.}~\bibnamefont {Ruud}}, \ and\
  \bibinfo {author} {\bibfnamefont {Y.}~\bibnamefont {Luo}},\ }\bibfield
  {title} {\enquote {\bibinfo {title} {Shape-dependent electronic excitations
  in metallic chains},}\ }\href {\doibase 10.1021/jp5000107} {\bibfield
  {journal} {\bibinfo  {journal} {Journal of Physical Chemistry C}\ }\textbf
  {\bibinfo {volume} {118}},\ \bibinfo {pages} {13059--13069} (\bibinfo {year}
  {2014})}\BibitemShut {NoStop}%
\bibitem [{\citenamefont {Guidez}\ and\ \citenamefont
  {Aikens}(2014{\natexlab{a}})}]{Guidez201411512}%
  \BibitemOpen
  \bibfield  {author} {\bibinfo {author} {\bibfnamefont {E.}~\bibnamefont
  {Guidez}}\ and\ \bibinfo {author} {\bibfnamefont {C.}~\bibnamefont
  {Aikens}},\ }\bibfield  {title} {\enquote {\bibinfo {title} {Quantum
  mechanical origin of the plasmon: From molecular systems to nanoparticles},}\
  }\href {\doibase 10.1039/c4nr02225d} {\bibfield  {journal} {\bibinfo
  {journal} {Nanoscale}\ }\textbf {\bibinfo {volume} {6}},\ \bibinfo {pages}
  {11512--11527} (\bibinfo {year} {2014}{\natexlab{a}})}\BibitemShut {NoStop}%
\bibitem [{\citenamefont {Conley}\ \emph {et~al.}(2019)\citenamefont {Conley},
  \citenamefont {Nayyar}, \citenamefont {Rossi}, \citenamefont {Kuisma},
  \citenamefont {Turkowski}, \citenamefont {Puska},\ and\ \citenamefont
  {Rahman}}]{Conley20195344}%
  \BibitemOpen
  \bibfield  {author} {\bibinfo {author} {\bibfnamefont {K.}~\bibnamefont
  {Conley}}, \bibinfo {author} {\bibfnamefont {N.}~\bibnamefont {Nayyar}},
  \bibinfo {author} {\bibfnamefont {T.}~\bibnamefont {Rossi}}, \bibinfo
  {author} {\bibfnamefont {M.}~\bibnamefont {Kuisma}}, \bibinfo {author}
  {\bibfnamefont {V.}~\bibnamefont {Turkowski}}, \bibinfo {author}
  {\bibfnamefont {M.}~\bibnamefont {Puska}}, \ and\ \bibinfo {author}
  {\bibfnamefont {T.}~\bibnamefont {Rahman}},\ }\bibfield  {title} {\enquote
  {\bibinfo {title} {Plasmon excitations in mixed metallic nanoarrays},}\
  }\href {\doibase 10.1021/acsnano.8b09826} {\bibfield  {journal} {\bibinfo
  {journal} {ACS Nano}\ }\textbf {\bibinfo {volume} {13}},\ \bibinfo {pages}
  {5344--5355} (\bibinfo {year} {2019})}\BibitemShut {NoStop}%
\bibitem [{\citenamefont {Peng}\ and\ \citenamefont
  {Kowalski}(2017)}]{Peng20174179}%
  \BibitemOpen
  \bibfield  {author} {\bibinfo {author} {\bibfnamefont {B.}~\bibnamefont
  {Peng}}\ and\ \bibinfo {author} {\bibfnamefont {K.}~\bibnamefont
  {Kowalski}},\ }\bibfield  {title} {\enquote {\bibinfo {title} {Highly
  efficient and scalable compound decomposition of two-electron integral tensor
  and its application in coupled cluster calculations},}\ }\href {\doibase
  10.1021/acs.jctc.7b00605} {\bibfield  {journal} {\bibinfo  {journal} {Journal
  of Chemical Theory and Computation}\ }\textbf {\bibinfo {volume} {13}},\
  \bibinfo {pages} {4179--4192} (\bibinfo {year} {2017})}\BibitemShut {NoStop}%
\bibitem [{\citenamefont {Epifanovsky}\ \emph {et~al.}(2013)\citenamefont
  {Epifanovsky}, \citenamefont {Zuev}, \citenamefont {Feng}, \citenamefont
  {Khistyaev}, \citenamefont {Shao},\ and\ \citenamefont
  {Krylov}}]{Epifanovsky2013}%
  \BibitemOpen
  \bibfield  {author} {\bibinfo {author} {\bibfnamefont {E.}~\bibnamefont
  {Epifanovsky}}, \bibinfo {author} {\bibfnamefont {D.}~\bibnamefont {Zuev}},
  \bibinfo {author} {\bibfnamefont {X.}~\bibnamefont {Feng}}, \bibinfo {author}
  {\bibfnamefont {K.}~\bibnamefont {Khistyaev}}, \bibinfo {author}
  {\bibfnamefont {Y.}~\bibnamefont {Shao}}, \ and\ \bibinfo {author}
  {\bibfnamefont {A.}~\bibnamefont {Krylov}},\ }\bibfield  {title} {\enquote
  {\bibinfo {title} {General implementation of the resolution-of-the-identity
  and cholesky representations of electron repulsion integrals within
  coupled-cluster and equation-of-motion methods: Theory and benchmarks},}\
  }\href {\doibase 10.1063/1.4820484} {\bibfield  {journal} {\bibinfo
  {journal} {Journal of Chemical Physics}\ }\textbf {\bibinfo {volume} {139}}
  (\bibinfo {year} {2013}),\ 10.1063/1.4820484}\BibitemShut {NoStop}%
\bibitem [{\citenamefont {Koch}, \citenamefont {Sánchez De~Merás},\ and\
  \citenamefont {Pedersen}(2003)}]{Koch20039481}%
  \BibitemOpen
  \bibfield  {author} {\bibinfo {author} {\bibfnamefont {H.}~\bibnamefont
  {Koch}}, \bibinfo {author} {\bibfnamefont {A.}~\bibnamefont {Sánchez
  De~Merás}}, \ and\ \bibinfo {author} {\bibfnamefont {T.}~\bibnamefont
  {Pedersen}},\ }\bibfield  {title} {\enquote {\bibinfo {title} {Reduced
  scaling in electronic structure calculations using cholesky
  decompositions},}\ }\href {\doibase 10.1063/1.1578621} {\bibfield  {journal}
  {\bibinfo  {journal} {Journal of Chemical Physics}\ }\textbf {\bibinfo
  {volume} {118}},\ \bibinfo {pages} {9481--9484} (\bibinfo {year}
  {2003})}\BibitemShut {NoStop}%
\bibitem [{\citenamefont {Beebe}\ and\ \citenamefont
  {Linderberg}(1977)}]{Beebe1977683}%
  \BibitemOpen
  \bibfield  {author} {\bibinfo {author} {\bibfnamefont {N.}~\bibnamefont
  {Beebe}}\ and\ \bibinfo {author} {\bibfnamefont {J.}~\bibnamefont
  {Linderberg}},\ }\bibfield  {title} {\enquote {\bibinfo {title}
  {Simplifications in the generation and transformation of two‐electron
  integrals in molecular calculations},}\ }\href {\doibase
  10.1002/qua.560120408} {\bibfield  {journal} {\bibinfo  {journal}
  {International Journal of Quantum Chemistry}\ }\textbf {\bibinfo {volume}
  {12}},\ \bibinfo {pages} {683--705} (\bibinfo {year} {1977})}\BibitemShut
  {NoStop}%
\bibitem [{\citenamefont {Krisiloff}\ \emph {et~al.}(2015)\citenamefont
  {Krisiloff}, \citenamefont {Krauter}, \citenamefont {Ricci},\ and\
  \citenamefont {Carter}}]{Krisiloff20155242}%
  \BibitemOpen
  \bibfield  {author} {\bibinfo {author} {\bibfnamefont {D.}~\bibnamefont
  {Krisiloff}}, \bibinfo {author} {\bibfnamefont {C.}~\bibnamefont {Krauter}},
  \bibinfo {author} {\bibfnamefont {F.}~\bibnamefont {Ricci}}, \ and\ \bibinfo
  {author} {\bibfnamefont {E.}~\bibnamefont {Carter}},\ }\bibfield  {title}
  {\enquote {\bibinfo {title} {Density fitting and cholesky decomposition of
  the two-electron integrals in local multireference configuration interaction
  theory},}\ }\href {\doibase 10.1021/acs.jctc.5b00762} {\bibfield  {journal}
  {\bibinfo  {journal} {Journal of Chemical Theory and Computation}\ }\textbf
  {\bibinfo {volume} {11}},\ \bibinfo {pages} {5242--5251} (\bibinfo {year}
  {2015})}\BibitemShut {NoStop}%
\bibitem [{\citenamefont {Bozkaya}\ and\ \citenamefont
  {Sherrill}(2017)}]{Bozkaya2017}%
  \BibitemOpen
  \bibfield  {author} {\bibinfo {author} {\bibfnamefont {U.}~\bibnamefont
  {Bozkaya}}\ and\ \bibinfo {author} {\bibfnamefont {C.}~\bibnamefont
  {Sherrill}},\ }\bibfield  {title} {\enquote {\bibinfo {title} {Analytic
  energy gradients for the coupled-cluster singles and doubles with
  perturbative triples method with the density-fitting approximation},}\ }\href
  {\doibase 10.1063/1.4994918} {\bibfield  {journal} {\bibinfo  {journal}
  {Journal of Chemical Physics}\ }\textbf {\bibinfo {volume} {147}} (\bibinfo
  {year} {2017}),\ 10.1063/1.4994918}\BibitemShut {NoStop}%
\bibitem [{\citenamefont {Wang}\ \emph {et~al.}(2016)\citenamefont {Wang},
  \citenamefont {Sokolov}, \citenamefont {Turney},\ and\ \citenamefont
  {Schaefer}}]{Wang20164833}%
  \BibitemOpen
  \bibfield  {author} {\bibinfo {author} {\bibfnamefont {X.}~\bibnamefont
  {Wang}}, \bibinfo {author} {\bibfnamefont {A.}~\bibnamefont {Sokolov}},
  \bibinfo {author} {\bibfnamefont {J.}~\bibnamefont {Turney}}, \ and\ \bibinfo
  {author} {\bibfnamefont {H.}~\bibnamefont {Schaefer}},\ }\bibfield  {title}
  {\enquote {\bibinfo {title} {Spin-adapted formulation and implementation of
  density cumulant functional theory with density-fitting approximation:
  Application to transition metal compounds},}\ }\href {\doibase
  10.1021/acs.jctc.6b00589} {\bibfield  {journal} {\bibinfo  {journal} {Journal
  of Chemical Theory and Computation}\ }\textbf {\bibinfo {volume} {12}},\
  \bibinfo {pages} {4833--4842} (\bibinfo {year} {2016})}\BibitemShut {NoStop}%
\bibitem [{\citenamefont {DePrince~III}\ \emph {et~al.}(2014)\citenamefont
  {DePrince~III}, \citenamefont {Kennedy}, \citenamefont {Sumpter},\ and\
  \citenamefont {Sherrill}}]{DePrinceIII2014844}%
  \BibitemOpen
  \bibfield  {author} {\bibinfo {author} {\bibfnamefont {A.}~\bibnamefont
  {DePrince~III}}, \bibinfo {author} {\bibfnamefont {M.}~\bibnamefont
  {Kennedy}}, \bibinfo {author} {\bibfnamefont {B.}~\bibnamefont {Sumpter}}, \
  and\ \bibinfo {author} {\bibfnamefont {C.}~\bibnamefont {Sherrill}},\
  }\bibfield  {title} {\enquote {\bibinfo {title} {Density-fitted singles and
  doubles coupled cluster on graphics processing units},}\ }\href {\doibase
  10.1080/00268976.2013.874599} {\bibfield  {journal} {\bibinfo  {journal}
  {Molecular Physics}\ }\textbf {\bibinfo {volume} {112}},\ \bibinfo {pages}
  {844--852} (\bibinfo {year} {2014})}\BibitemShut {NoStop}%
\bibitem [{\citenamefont {Neese}(2003)}]{Neese20031740}%
  \BibitemOpen
  \bibfield  {author} {\bibinfo {author} {\bibfnamefont {F.}~\bibnamefont
  {Neese}},\ }\bibfield  {title} {\enquote {\bibinfo {title} {An improvement of
  the resolution of the identity approximation for the formation of the coulomb
  matrix},}\ }\href {\doibase 10.1002/jcc.10318} {\bibfield  {journal}
  {\bibinfo  {journal} {Journal of Computational Chemistry}\ }\textbf {\bibinfo
  {volume} {24}},\ \bibinfo {pages} {1740--1747} (\bibinfo {year}
  {2003})}\BibitemShut {NoStop}%
\bibitem [{\citenamefont {Hohenstein}, \citenamefont {Parrish},\ and\
  \citenamefont {Martínez}(2012)}]{Hohenstein20121085}%
  \BibitemOpen
  \bibfield  {author} {\bibinfo {author} {\bibfnamefont {E.}~\bibnamefont
  {Hohenstein}}, \bibinfo {author} {\bibfnamefont {R.}~\bibnamefont {Parrish}},
  \ and\ \bibinfo {author} {\bibfnamefont {T.}~\bibnamefont {Martínez}},\
  }\bibfield  {title} {\enquote {\bibinfo {title} {Tensor hypercontraction
  density fitting. i. quartic scaling second- and third-order møller-plesset
  perturbation theory},}\ }\href {\doibase 10.1063/1.4732310} {\bibfield
  {journal} {\bibinfo  {journal} {Journal of Chemical Physics}\ }\textbf
  {\bibinfo {volume} {137}},\ \bibinfo {pages} {1085} (\bibinfo {year}
  {2012})}\BibitemShut {NoStop}%
\bibitem [{\citenamefont {Parrish}\ \emph {et~al.}(2012)\citenamefont
  {Parrish}, \citenamefont {Hohenstein}, \citenamefont {Martínez},\ and\
  \citenamefont {Sherrill}}]{Parrish2012}%
  \BibitemOpen
  \bibfield  {author} {\bibinfo {author} {\bibfnamefont {R.}~\bibnamefont
  {Parrish}}, \bibinfo {author} {\bibfnamefont {E.}~\bibnamefont {Hohenstein}},
  \bibinfo {author} {\bibfnamefont {T.}~\bibnamefont {Martínez}}, \ and\
  \bibinfo {author} {\bibfnamefont {C.}~\bibnamefont {Sherrill}},\ }\bibfield
  {title} {\enquote {\bibinfo {title} {Tensor hypercontraction. ii.
  least-squares renormalization},}\ }\href {\doibase 10.1063/1.4768233}
  {\bibfield  {journal} {\bibinfo  {journal} {Journal of Chemical Physics}\
  }\textbf {\bibinfo {volume} {137}} (\bibinfo {year} {2012}),\
  10.1063/1.4768233}\BibitemShut {NoStop}%
\bibitem [{\citenamefont {Hohenstein}\ \emph {et~al.}(2012)\citenamefont
  {Hohenstein}, \citenamefont {Parrish}, \citenamefont {Sherrill},\ and\
  \citenamefont {Martínez}}]{Hohenstein2012}%
  \BibitemOpen
  \bibfield  {author} {\bibinfo {author} {\bibfnamefont {E.}~\bibnamefont
  {Hohenstein}}, \bibinfo {author} {\bibfnamefont {R.}~\bibnamefont {Parrish}},
  \bibinfo {author} {\bibfnamefont {C.}~\bibnamefont {Sherrill}}, \ and\
  \bibinfo {author} {\bibfnamefont {T.}~\bibnamefont {Martínez}},\ }\bibfield
  {title} {\enquote {\bibinfo {title} {Communication: Tensor hypercontraction.
  iii. least-squares tensor hypercontraction for the determination of
  correlated wavefunctions},}\ }\href {\doibase 10.1063/1.4768241} {\bibfield
  {journal} {\bibinfo  {journal} {Journal of Chemical Physics}\ }\textbf
  {\bibinfo {volume} {137}} (\bibinfo {year} {2012}),\
  10.1063/1.4768241}\BibitemShut {NoStop}%
\bibitem [{\citenamefont {Dreuw}\ and\ \citenamefont
  {Head-Gordon}(2005)}]{Dreuw20054009}%
  \BibitemOpen
  \bibfield  {author} {\bibinfo {author} {\bibfnamefont {A.}~\bibnamefont
  {Dreuw}}\ and\ \bibinfo {author} {\bibfnamefont {M.}~\bibnamefont
  {Head-Gordon}},\ }\bibfield  {title} {\enquote {\bibinfo {title}
  {Single-reference ab initio methods for the calculation of excited states of
  large molecules},}\ }\href {\doibase 10.1021/cr0505627} {\bibfield  {journal}
  {\bibinfo  {journal} {Chemical Reviews}\ }\textbf {\bibinfo {volume} {105}},\
  \bibinfo {pages} {4009--4037} (\bibinfo {year} {2005})}\BibitemShut {NoStop}%
\bibitem [{\citenamefont {Casida}\ and\ \citenamefont
  {Huix-Rotllant}(2012)}]{Casida2012287}%
  \BibitemOpen
  \bibfield  {author} {\bibinfo {author} {\bibfnamefont {M.}~\bibnamefont
  {Casida}}\ and\ \bibinfo {author} {\bibfnamefont {M.}~\bibnamefont
  {Huix-Rotllant}},\ }\bibfield  {title} {\enquote {\bibinfo {title} {Progress
  in time-dependent density-functional theory},}\ }\href {\doibase
  10.1146/annurev-physchem-032511-143803} {\bibfield  {journal} {\bibinfo
  {journal} {Annual Review of Physical Chemistry}\ }\textbf {\bibinfo {volume}
  {63}},\ \bibinfo {pages} {287--323} (\bibinfo {year} {2012})}\BibitemShut
  {NoStop}%
\bibitem [{\citenamefont {Shavitt}(2009)}]{ShavittBartlett2009}%
  \BibitemOpen
  \bibfield  {author} {\bibinfo {author} {\bibfnamefont {I.~B. R.~J.}\
  \bibnamefont {Shavitt}},\ }\href@noop {} {\emph {\bibinfo {title} {Many-Body
  Methods in Chemistry and Physics}}}\ (\bibinfo {year} {2009})\BibitemShut
  {NoStop}%
\bibitem [{\citenamefont {Fetter}\ and\ \citenamefont
  {Walecka}(1971)}]{Fetter1971}%
  \BibitemOpen
  \bibfield  {author} {\bibinfo {author} {\bibfnamefont {A.}~\bibnamefont
  {Fetter}}\ and\ \bibinfo {author} {\bibfnamefont {J.}~\bibnamefont
  {Walecka}},\ }\href@noop {} {\bibfield  {journal} {\bibinfo  {journal}
  {Quantum Theory of Many-Particle Systems}\ } (\bibinfo {year}
  {1971})}\BibitemShut {NoStop}%
\bibitem [{\citenamefont {Onida}, \citenamefont {Reining},\ and\ \citenamefont
  {Rubio}(2002)}]{Onida2002601}%
  \BibitemOpen
  \bibfield  {author} {\bibinfo {author} {\bibfnamefont {G.}~\bibnamefont
  {Onida}}, \bibinfo {author} {\bibfnamefont {L.}~\bibnamefont {Reining}}, \
  and\ \bibinfo {author} {\bibfnamefont {A.}~\bibnamefont {Rubio}},\ }\bibfield
   {title} {\enquote {\bibinfo {title} {Electronic excitations:
  Density-functional versus many-body green's-function approaches},}\ }\href
  {\doibase 10.1103/RevModPhys.74.601} {\bibfield  {journal} {\bibinfo
  {journal} {Reviews of Modern Physics}\ }\textbf {\bibinfo {volume} {74}},\
  \bibinfo {pages} {601--659} (\bibinfo {year} {2002})}\BibitemShut {NoStop}%
\bibitem [{\citenamefont {Blase}, \citenamefont {Duchemin},\ and\ \citenamefont
  {Jacquemin}(2018)}]{Blase20181022}%
  \BibitemOpen
  \bibfield  {author} {\bibinfo {author} {\bibfnamefont {X.}~\bibnamefont
  {Blase}}, \bibinfo {author} {\bibfnamefont {I.}~\bibnamefont {Duchemin}}, \
  and\ \bibinfo {author} {\bibfnamefont {D.}~\bibnamefont {Jacquemin}},\
  }\bibfield  {title} {\enquote {\bibinfo {title} {The bethe-salpeter equation
  in chemistry: Relations with td-dft, applications and challenges},}\ }\href
  {\doibase 10.1039/c7cs00049a} {\bibfield  {journal} {\bibinfo  {journal}
  {Chemical Society Reviews}\ }\textbf {\bibinfo {volume} {47}},\ \bibinfo
  {pages} {1022--1043} (\bibinfo {year} {2018})}\BibitemShut {NoStop}%
\bibitem [{\citenamefont {Govoni}\ and\ \citenamefont
  {Galli}(2015)}]{Govoni20152680}%
  \BibitemOpen
  \bibfield  {author} {\bibinfo {author} {\bibfnamefont {M.}~\bibnamefont
  {Govoni}}\ and\ \bibinfo {author} {\bibfnamefont {G.}~\bibnamefont {Galli}},\
  }\bibfield  {title} {\enquote {\bibinfo {title} {Large scale gw
  calculations},}\ }\href {\doibase 10.1021/ct500958p} {\bibfield  {journal}
  {\bibinfo  {journal} {J. Chem. Theory Comput.}\ }\textbf {\bibinfo {volume}
  {11}},\ \bibinfo {pages} {2680--2696} (\bibinfo {year} {2015})}\BibitemShut
  {NoStop}%
\bibitem [{\citenamefont {Weerawardene}\ and\ \citenamefont
  {Aikens}(2018)}]{Weerawardene201827}%
  \BibitemOpen
  \bibfield  {author} {\bibinfo {author} {\bibfnamefont {K.}~\bibnamefont
  {Weerawardene}}\ and\ \bibinfo {author} {\bibfnamefont {C.}~\bibnamefont
  {Aikens}},\ }\bibfield  {title} {\enquote {\bibinfo {title} {Comparison and
  convergence of optical absorption spectra of noble metal nanoparticles
  computed using linear-response and real-time time-dependent density
  functional theories},}\ }\href {\doibase 10.1016/j.comptc.2018.11.005}
  {\bibfield  {journal} {\bibinfo  {journal} {Computational and Theoretical
  Chemistry}\ }\textbf {\bibinfo {volume} {1146}},\ \bibinfo {pages} {27--36}
  (\bibinfo {year} {2018})}\BibitemShut {NoStop}%
\bibitem [{\citenamefont {Ding}\ \emph {et~al.}(2014)\citenamefont {Ding},
  \citenamefont {Guidez}, \citenamefont {Aikens},\ and\ \citenamefont
  {Li}}]{Ding2014}%
  \BibitemOpen
  \bibfield  {author} {\bibinfo {author} {\bibfnamefont {F.}~\bibnamefont
  {Ding}}, \bibinfo {author} {\bibfnamefont {E.}~\bibnamefont {Guidez}},
  \bibinfo {author} {\bibfnamefont {C.}~\bibnamefont {Aikens}}, \ and\ \bibinfo
  {author} {\bibfnamefont {X.}~\bibnamefont {Li}},\ }\bibfield  {title}
  {\enquote {\bibinfo {title} {Quantum coherent plasmon in silver nanowires: A
  real-time tddft study},}\ }\href {\doibase 10.1063/1.4884388} {\bibfield
  {journal} {\bibinfo  {journal} {Journal of Chemical Physics}\ }\textbf
  {\bibinfo {volume} {140}} (\bibinfo {year} {2014}),\
  10.1063/1.4884388}\BibitemShut {NoStop}%
\bibitem [{\citenamefont {Peng}\ \emph
  {et~al.}(2015{\natexlab{a}})\citenamefont {Peng}, \citenamefont {Lingerfelt},
  \citenamefont {Ding}, \citenamefont {Aikens},\ and\ \citenamefont
  {Li}}]{Peng20156421}%
  \BibitemOpen
  \bibfield  {author} {\bibinfo {author} {\bibfnamefont {B.}~\bibnamefont
  {Peng}}, \bibinfo {author} {\bibfnamefont {D.}~\bibnamefont {Lingerfelt}},
  \bibinfo {author} {\bibfnamefont {F.}~\bibnamefont {Ding}}, \bibinfo {author}
  {\bibfnamefont {C.}~\bibnamefont {Aikens}}, \ and\ \bibinfo {author}
  {\bibfnamefont {X.}~\bibnamefont {Li}},\ }\bibfield  {title} {\enquote
  {\bibinfo {title} {Real-time tddft studies of exciton decay and transfer in
  silver nanowire arrays},}\ }\href {\doibase 10.1021/acs.jpcc.5b00263}
  {\bibfield  {journal} {\bibinfo  {journal} {Journal of Physical Chemistry C}\
  }\textbf {\bibinfo {volume} {119}},\ \bibinfo {pages} {6421--6427} (\bibinfo
  {year} {2015}{\natexlab{a}})}\BibitemShut {NoStop}%
\bibitem [{\citenamefont {Gao}, \citenamefont {Ruud},\ and\ \citenamefont
  {Luo}(2012)}]{Gao2012}%
  \BibitemOpen
  \bibfield  {author} {\bibinfo {author} {\bibfnamefont {B.}~\bibnamefont
  {Gao}}, \bibinfo {author} {\bibfnamefont {K.}~\bibnamefont {Ruud}}, \ and\
  \bibinfo {author} {\bibfnamefont {Y.}~\bibnamefont {Luo}},\ }\bibfield
  {title} {\enquote {\bibinfo {title} {Plasmon resonances in linear noble-metal
  chains},}\ }\href {\doibase 10.1063/1.4766360} {\bibfield  {journal}
  {\bibinfo  {journal} {Journal of Chemical Physics}\ }\textbf {\bibinfo
  {volume} {137}} (\bibinfo {year} {2012}),\ 10.1063/1.4766360}\BibitemShut
  {NoStop}%
\bibitem [{\citenamefont {Senanayake}\ \emph {et~al.}(2019)\citenamefont
  {Senanayake}, \citenamefont {Lingerfelt}, \citenamefont {Kuda-Singappulige},
  \citenamefont {Li},\ and\ \citenamefont {Aikens}}]{Senanayake201914734}%
  \BibitemOpen
  \bibfield  {author} {\bibinfo {author} {\bibfnamefont {R.}~\bibnamefont
  {Senanayake}}, \bibinfo {author} {\bibfnamefont {D.}~\bibnamefont
  {Lingerfelt}}, \bibinfo {author} {\bibfnamefont {G.}~\bibnamefont
  {Kuda-Singappulige}}, \bibinfo {author} {\bibfnamefont {X.}~\bibnamefont
  {Li}}, \ and\ \bibinfo {author} {\bibfnamefont {C.}~\bibnamefont {Aikens}},\
  }\bibfield  {title} {\enquote {\bibinfo {title} {Real-time tddft
  investigation of optical absorption in gold nanowires},}\ }\href {\doibase
  10.1021/acs.jpcc.9b00296} {\bibfield  {journal} {\bibinfo  {journal} {Journal
  of Physical Chemistry C}\ }\textbf {\bibinfo {volume} {123}},\ \bibinfo
  {pages} {14734--14745} (\bibinfo {year} {2019})}\BibitemShut {NoStop}%
\bibitem [{\citenamefont {Piccini}\ \emph {et~al.}(2013)\citenamefont
  {Piccini}, \citenamefont {Havenith}, \citenamefont {Broer},\ and\
  \citenamefont {Stener}}]{Piccini201317196}%
  \BibitemOpen
  \bibfield  {author} {\bibinfo {author} {\bibfnamefont {G.}~\bibnamefont
  {Piccini}}, \bibinfo {author} {\bibfnamefont {R.}~\bibnamefont {Havenith}},
  \bibinfo {author} {\bibfnamefont {R.}~\bibnamefont {Broer}}, \ and\ \bibinfo
  {author} {\bibfnamefont {M.}~\bibnamefont {Stener}},\ }\bibfield  {title}
  {\enquote {\bibinfo {title} {Gold nanowires: A time-dependent density
  functional assessment of plasmonic behavior},}\ }\href {\doibase
  10.1021/jp405769e} {\bibfield  {journal} {\bibinfo  {journal} {Journal of
  Physical Chemistry C}\ }\textbf {\bibinfo {volume} {117}},\ \bibinfo {pages}
  {17196--17204} (\bibinfo {year} {2013})}\BibitemShut {NoStop}%
\bibitem [{\citenamefont {Barcaro}\ \emph {et~al.}(2014)\citenamefont
  {Barcaro}, \citenamefont {Sementa}, \citenamefont {Fortunelli},\ and\
  \citenamefont {Stener}}]{Barcaro201412450}%
  \BibitemOpen
  \bibfield  {author} {\bibinfo {author} {\bibfnamefont {G.}~\bibnamefont
  {Barcaro}}, \bibinfo {author} {\bibfnamefont {L.}~\bibnamefont {Sementa}},
  \bibinfo {author} {\bibfnamefont {A.}~\bibnamefont {Fortunelli}}, \ and\
  \bibinfo {author} {\bibfnamefont {M.}~\bibnamefont {Stener}},\ }\bibfield
  {title} {\enquote {\bibinfo {title} {Optical properties of silver nanoshells
  from time-dependent density functional theory calculations},}\ }\href
  {\doibase 10.1021/jp5016565} {\bibfield  {journal} {\bibinfo  {journal}
  {Journal of Physical Chemistry C}\ }\textbf {\bibinfo {volume} {118}},\
  \bibinfo {pages} {12450--12458} (\bibinfo {year} {2014})}\BibitemShut
  {NoStop}%
\bibitem [{\citenamefont {Baseggio}\ \emph {et~al.}(2016)\citenamefont
  {Baseggio}, \citenamefont {De~Vetta}, \citenamefont {Fronzoni}, \citenamefont
  {Stener}, \citenamefont {Sementa}, \citenamefont {Fortunelli},\ and\
  \citenamefont {Calzolari}}]{Baseggio201612773}%
  \BibitemOpen
  \bibfield  {author} {\bibinfo {author} {\bibfnamefont {O.}~\bibnamefont
  {Baseggio}}, \bibinfo {author} {\bibfnamefont {M.}~\bibnamefont {De~Vetta}},
  \bibinfo {author} {\bibfnamefont {G.}~\bibnamefont {Fronzoni}}, \bibinfo
  {author} {\bibfnamefont {M.}~\bibnamefont {Stener}}, \bibinfo {author}
  {\bibfnamefont {L.}~\bibnamefont {Sementa}}, \bibinfo {author} {\bibfnamefont
  {A.}~\bibnamefont {Fortunelli}}, \ and\ \bibinfo {author} {\bibfnamefont
  {A.}~\bibnamefont {Calzolari}},\ }\bibfield  {title} {\enquote {\bibinfo
  {title} {Photoabsorption of icosahedral noble metal clusters: An efficient
  tddft approach to large-scale systems},}\ }\href {\doibase
  10.1021/acs.jpcc.6b04709} {\bibfield  {journal} {\bibinfo  {journal} {Journal
  of Physical Chemistry C}\ }\textbf {\bibinfo {volume} {120}},\ \bibinfo
  {pages} {12773--12782} (\bibinfo {year} {2016})}\BibitemShut {NoStop}%
\bibitem [{\citenamefont {Ma}, \citenamefont {Wang},\ and\ \citenamefont
  {Pei}(2016)}]{Ma201617044}%
  \BibitemOpen
  \bibfield  {author} {\bibinfo {author} {\bibfnamefont {Z.}~\bibnamefont
  {Ma}}, \bibinfo {author} {\bibfnamefont {P.}~\bibnamefont {Wang}}, \ and\
  \bibinfo {author} {\bibfnamefont {Y.}~\bibnamefont {Pei}},\ }\bibfield
  {title} {\enquote {\bibinfo {title} {Geometric structure, electronic
  structure and optical absorption properties of one-dimensional
  thiolate-protected gold clusters containing a quasi-face-centered-cubic
  (quasi-fcc) au-core: A density-functional theoretical study},}\ }\href
  {\doibase 10.1039/c6nr04998b} {\bibfield  {journal} {\bibinfo  {journal}
  {Nanoscale}\ }\textbf {\bibinfo {volume} {8}},\ \bibinfo {pages}
  {17044--17054} (\bibinfo {year} {2016})}\BibitemShut {NoStop}%
\bibitem [{\citenamefont {Zhang}\ and\ \citenamefont
  {Zhang}(2014)}]{Zhang2014635}%
  \BibitemOpen
  \bibfield  {author} {\bibinfo {author} {\bibfnamefont {K.}~\bibnamefont
  {Zhang}}\ and\ \bibinfo {author} {\bibfnamefont {H.}~\bibnamefont {Zhang}},\
  }\bibfield  {title} {\enquote {\bibinfo {title} {Plasmon coupling in gold
  nanotube assemblies: Insight from a time-dependent density functional theory
  (tddft) calculation},}\ }\href {\doibase 10.1021/jp410056u} {\bibfield
  {journal} {\bibinfo  {journal} {Journal of Physical Chemistry C}\ }\textbf
  {\bibinfo {volume} {118}},\ \bibinfo {pages} {635--641} (\bibinfo {year}
  {2014})}\BibitemShut {NoStop}%
\bibitem [{\citenamefont {Fernando}\ \emph {et~al.}(2015)\citenamefont
  {Fernando}, \citenamefont {Weerawardene}, \citenamefont {Karimova},\ and\
  \citenamefont {Aikens}}]{Fernando20156112}%
  \BibitemOpen
  \bibfield  {author} {\bibinfo {author} {\bibfnamefont {A.}~\bibnamefont
  {Fernando}}, \bibinfo {author} {\bibfnamefont {K.}~\bibnamefont
  {Weerawardene}}, \bibinfo {author} {\bibfnamefont {N.}~\bibnamefont
  {Karimova}}, \ and\ \bibinfo {author} {\bibfnamefont {C.}~\bibnamefont
  {Aikens}},\ }\bibfield  {title} {\enquote {\bibinfo {title} {Quantum
  mechanical studies of large metal, metal oxide, and metal chalcogenide
  nanoparticles and clusters},}\ }\href {\doibase 10.1021/cr500506r} {\bibfield
   {journal} {\bibinfo  {journal} {Chemical Reviews}\ }\textbf {\bibinfo
  {volume} {115}},\ \bibinfo {pages} {6112--6216} (\bibinfo {year}
  {2015})}\BibitemShut {NoStop}%
\bibitem [{\citenamefont {Guidez}\ and\ \citenamefont
  {Aikens}(2014{\natexlab{b}})}]{Guidez201415501}%
  \BibitemOpen
  \bibfield  {author} {\bibinfo {author} {\bibfnamefont {E.}~\bibnamefont
  {Guidez}}\ and\ \bibinfo {author} {\bibfnamefont {C.}~\bibnamefont
  {Aikens}},\ }\bibfield  {title} {\enquote {\bibinfo {title} {Plasmon
  resonance analysis with configuration interaction},}\ }\href {\doibase
  10.1039/c4cp01365d} {\bibfield  {journal} {\bibinfo  {journal} {Physical
  Chemistry Chemical Physics}\ }\textbf {\bibinfo {volume} {16}},\ \bibinfo
  {pages} {15501--15509} (\bibinfo {year} {2014}{\natexlab{b}})}\BibitemShut
  {NoStop}%
\bibitem [{\citenamefont {Bae}\ and\ \citenamefont
  {Aikens}(2012{\natexlab{b}})}]{Bae20128260}%
  \BibitemOpen
  \bibfield  {author} {\bibinfo {author} {\bibfnamefont {G.-T.}\ \bibnamefont
  {Bae}}\ and\ \bibinfo {author} {\bibfnamefont {C.}~\bibnamefont {Aikens}},\
  }\bibfield  {title} {\enquote {\bibinfo {title} {Tddft and cis studies of
  optical properties of dimers of silver tetrahedra},}\ }\href {\doibase
  10.1021/jp305330e} {\bibfield  {journal} {\bibinfo  {journal} {Journal of
  Physical Chemistry A}\ }\textbf {\bibinfo {volume} {116}},\ \bibinfo {pages}
  {8260--8269} (\bibinfo {year} {2012}{\natexlab{b}})}\BibitemShut {NoStop}%
\bibitem [{\citenamefont {Fales}, \citenamefont {Hohenstein},\ and\
  \citenamefont {Levine}(2017)}]{Fales20174162}%
  \BibitemOpen
  \bibfield  {author} {\bibinfo {author} {\bibfnamefont {B.}~\bibnamefont
  {Fales}}, \bibinfo {author} {\bibfnamefont {E.}~\bibnamefont {Hohenstein}}, \
  and\ \bibinfo {author} {\bibfnamefont {B.}~\bibnamefont {Levine}},\
  }\bibfield  {title} {\enquote {\bibinfo {title} {Robust and efficient spin
  purification for determinantal configuration interaction},}\ }\href {\doibase
  10.1021/acs.jctc.7b00466} {\bibfield  {journal} {\bibinfo  {journal} {Journal
  of Chemical Theory and Computation}\ }\textbf {\bibinfo {volume} {13}},\
  \bibinfo {pages} {4162--4172} (\bibinfo {year} {2017})}\BibitemShut {NoStop}%
\bibitem [{\citenamefont {Bona\v{c}i\'{c}-Koutecky}, \citenamefont {Veyret},\
  and\ \citenamefont {Mitri\'{c}}(2001)}]{Koutecky200110450}%
  \BibitemOpen
  \bibfield  {author} {\bibinfo {author} {\bibfnamefont {V.}~\bibnamefont
  {Bona\v{c}i\'{c}-Koutecky}}, \bibinfo {author} {\bibfnamefont
  {V.}~\bibnamefont {Veyret}}, \ and\ \bibinfo {author} {\bibfnamefont
  {R.}~\bibnamefont {Mitri\'{c}}},\ }\bibfield  {title} {\enquote {\bibinfo
  {title} {Ab initio study of the absorption spectra of agn (n=5-8)
  clusters},}\ }\href {\doibase 10.1063/1.1415077} {\bibfield  {journal}
  {\bibinfo  {journal} {Journal of Chemical Physics}\ }\textbf {\bibinfo
  {volume} {115}},\ \bibinfo {pages} {10450--10460} (\bibinfo {year}
  {2001})}\BibitemShut {NoStop}%
\bibitem [{\citenamefont {McLaughlin}\ and\ \citenamefont
  {Chakraborty}(2020)}]{doi:10.1021/acs.jctc.9b01238}%
  \BibitemOpen
  \bibfield  {author} {\bibinfo {author} {\bibfnamefont {P.~F.}\ \bibnamefont
  {McLaughlin}}\ and\ \bibinfo {author} {\bibfnamefont {A.}~\bibnamefont
  {Chakraborty}},\ }\bibfield  {title} {\enquote {\bibinfo {title} {Compact
  real-space representation of excited states using frequency-dependent
  explicitly-correlated electron-hole interaction kernel},}\ }\href {\doibase
  10.1021/acs.jctc.9b01238} {\bibfield  {journal} {\bibinfo  {journal} {Journal
  of Chemical Theory and Computation}\ }\textbf {\bibinfo {volume} {ASAP}},\
  \bibinfo {pages} {ASAP} (\bibinfo {year} {2020})}\BibitemShut {NoStop}%
\bibitem [{\citenamefont {Goings}\ and\ \citenamefont {Li}(2016)}]{Goings2016}%
  \BibitemOpen
  \bibfield  {author} {\bibinfo {author} {\bibfnamefont {J.}~\bibnamefont
  {Goings}}\ and\ \bibinfo {author} {\bibfnamefont {X.}~\bibnamefont {Li}},\
  }\bibfield  {title} {\enquote {\bibinfo {title} {An atomic orbital based
  real-time time-dependent density functional theory for computing electronic
  circular dichroism band spectra},}\ }\href {\doibase 10.1063/1.4953668}
  {\bibfield  {journal} {\bibinfo  {journal} {Journal of Chemical Physics}\
  }\textbf {\bibinfo {volume} {144}} (\bibinfo {year} {2016}),\
  10.1063/1.4953668}\BibitemShut {NoStop}%
\bibitem [{\citenamefont {Liang}\ \emph {et~al.}(2011)\citenamefont {Liang},
  \citenamefont {Fischer}, \citenamefont {Frisch},\ and\ \citenamefont
  {Li}}]{Liang20113540}%
  \BibitemOpen
  \bibfield  {author} {\bibinfo {author} {\bibfnamefont {W.}~\bibnamefont
  {Liang}}, \bibinfo {author} {\bibfnamefont {S.}~\bibnamefont {Fischer}},
  \bibinfo {author} {\bibfnamefont {M.}~\bibnamefont {Frisch}}, \ and\ \bibinfo
  {author} {\bibfnamefont {X.}~\bibnamefont {Li}},\ }\bibfield  {title}
  {\enquote {\bibinfo {title} {Energy-specific linear response tdhf/tddft for
  calculating high-energy excited states},}\ }\href {\doibase
  10.1021/ct200485x} {\bibfield  {journal} {\bibinfo  {journal} {Journal of
  Chemical Theory and Computation}\ }\textbf {\bibinfo {volume} {7}},\ \bibinfo
  {pages} {3540--3547} (\bibinfo {year} {2011})}\BibitemShut {NoStop}%
\bibitem [{\citenamefont {Peng}\ \emph
  {et~al.}(2015{\natexlab{b}})\citenamefont {Peng}, \citenamefont {Lestrange},
  \citenamefont {Goings}, \citenamefont {Caricato},\ and\ \citenamefont
  {Li}}]{Peng20154146}%
  \BibitemOpen
  \bibfield  {author} {\bibinfo {author} {\bibfnamefont {B.}~\bibnamefont
  {Peng}}, \bibinfo {author} {\bibfnamefont {P.}~\bibnamefont {Lestrange}},
  \bibinfo {author} {\bibfnamefont {J.}~\bibnamefont {Goings}}, \bibinfo
  {author} {\bibfnamefont {M.}~\bibnamefont {Caricato}}, \ and\ \bibinfo
  {author} {\bibfnamefont {X.}~\bibnamefont {Li}},\ }\bibfield  {title}
  {\enquote {\bibinfo {title} {Energy-specific equation-of-motion
  coupled-cluster methods for high-energy excited states: Application to k-edge
  x-ray absorption spectroscopy},}\ }\href {\doibase 10.1021/acs.jctc.5b00459}
  {\bibfield  {journal} {\bibinfo  {journal} {Journal of Chemical Theory and
  Computation}\ }\textbf {\bibinfo {volume} {11}},\ \bibinfo {pages}
  {4146--4153} (\bibinfo {year} {2015}{\natexlab{b}})}\BibitemShut {NoStop}%
\bibitem [{\citenamefont {Zhu}, \citenamefont {Hybertsen},\ and\ \citenamefont
  {Littlewood}(1996)}]{Zhu199613575}%
  \BibitemOpen
  \bibfield  {author} {\bibinfo {author} {\bibfnamefont {X.}~\bibnamefont
  {Zhu}}, \bibinfo {author} {\bibfnamefont {M.}~\bibnamefont {Hybertsen}}, \
  and\ \bibinfo {author} {\bibfnamefont {P.}~\bibnamefont {Littlewood}},\
  }\bibfield  {title} {\enquote {\bibinfo {title} {Electron-hole system
  revisited: A variational quantum monte carlo study},}\ }\href {\doibase
  10.1103/PhysRevB.54.13575} {\bibfield  {journal} {\bibinfo  {journal}
  {Physical Review B - Condensed Matter and Materials Physics}\ }\textbf
  {\bibinfo {volume} {54}},\ \bibinfo {pages} {13575--13580} (\bibinfo {year}
  {1996})}\BibitemShut {NoStop}%
\bibitem [{\citenamefont {Ulrike}(1996)}]{Woggon1996}%
  \BibitemOpen
  \bibfield  {author} {\bibinfo {author} {\bibfnamefont {W.}~\bibnamefont
  {Ulrike}},\ }\href@noop {} {\emph {\bibinfo {title} {Optical Properties of
  Semiconductor Quantum Dots}}}\ (\bibinfo  {publisher} {Springer},\ \bibinfo
  {address} {Berlin, Heidelberg},\ \bibinfo {year} {1996})\BibitemShut
  {NoStop}%
\bibitem [{\citenamefont {Mattuck}(1976)}]{Mattuck1976}%
  \BibitemOpen
  \bibfield  {author} {\bibinfo {author} {\bibfnamefont {R.}~\bibnamefont
  {Mattuck}},\ }\href@noop {} {\emph {\bibinfo {title} {A Guide to Feynman
  Diagrams in the Many-Body Problem}}}\ (\bibinfo  {publisher} {Dover
  Publications},\ \bibinfo {year} {1976})\BibitemShut {NoStop}%
\bibitem [{\citenamefont {Elward}, \citenamefont {Thallinger},\ and\
  \citenamefont {Chakraborty}(2012)}]{Elward2012}%
  \BibitemOpen
  \bibfield  {author} {\bibinfo {author} {\bibfnamefont {J.}~\bibnamefont
  {Elward}}, \bibinfo {author} {\bibfnamefont {B.}~\bibnamefont {Thallinger}},
  \ and\ \bibinfo {author} {\bibfnamefont {A.}~\bibnamefont {Chakraborty}},\
  }\bibfield  {title} {\enquote {\bibinfo {title} {Calculation of electron-hole
  recombination probability using explicitly correlated hartree-fock method},}\
  }\href {\doibase 10.1063/1.3693765} {\bibfield  {journal} {\bibinfo
  {journal} {J. Chem. Phys.}\ }\textbf {\bibinfo {volume} {136}},\ \bibinfo
  {pages} {182--186} (\bibinfo {year} {2012})}\BibitemShut {NoStop}%
\bibitem [{\citenamefont {Ellis}, \citenamefont {Aggarwal},\ and\ \citenamefont
  {Chakraborty}(2016)}]{Ellis2016188}%
  \BibitemOpen
  \bibfield  {author} {\bibinfo {author} {\bibfnamefont {B.}~\bibnamefont
  {Ellis}}, \bibinfo {author} {\bibfnamefont {S.}~\bibnamefont {Aggarwal}}, \
  and\ \bibinfo {author} {\bibfnamefont {A.}~\bibnamefont {Chakraborty}},\
  }\bibfield  {title} {\enquote {\bibinfo {title} {Development of the
  multicomponent coupled-cluster theory for investigation of multiexcitonic
  interactions},}\ }\href {\doibase 10.1021/acs.jctc.5b00879} {\bibfield
  {journal} {\bibinfo  {journal} {Journal of Chemical Theory and Computation}\
  }\textbf {\bibinfo {volume} {12}},\ \bibinfo {pages} {188--200} (\bibinfo
  {year} {2016})}\BibitemShut {NoStop}%
\bibitem [{\citenamefont {Ellis}\ and\ \citenamefont
  {Chakraborty}(2017)}]{Ellis20171291}%
  \BibitemOpen
  \bibfield  {author} {\bibinfo {author} {\bibfnamefont {B.}~\bibnamefont
  {Ellis}}\ and\ \bibinfo {author} {\bibfnamefont {A.}~\bibnamefont
  {Chakraborty}},\ }\bibfield  {title} {\enquote {\bibinfo {title}
  {Investigation of many-body correlation in biexcitonic systems using
  electron-hole multicomponent coupled-cluster theory},}\ }\href {\doibase
  10.1021/acs.jpcc.6b09443} {\bibfield  {journal} {\bibinfo  {journal} {Journal
  of Physical Chemistry C}\ }\textbf {\bibinfo {volume} {121}},\ \bibinfo
  {pages} {1291--1298} (\bibinfo {year} {2017})}\BibitemShut {NoStop}%
\bibitem [{\citenamefont {Bayne}\ \emph {et~al.}(2018)\citenamefont {Bayne},
  \citenamefont {Scher}, \citenamefont {Ellis},\ and\ \citenamefont
  {Chakraborty}}]{Bayne20183656}%
  \BibitemOpen
  \bibfield  {author} {\bibinfo {author} {\bibfnamefont {M.}~\bibnamefont
  {Bayne}}, \bibinfo {author} {\bibfnamefont {J.}~\bibnamefont {Scher}},
  \bibinfo {author} {\bibfnamefont {B.}~\bibnamefont {Ellis}}, \ and\ \bibinfo
  {author} {\bibfnamefont {A.}~\bibnamefont {Chakraborty}},\ }\bibfield
  {title} {\enquote {\bibinfo {title} {Linked-cluster formulation of
  electron-hole interaction kernel in real-space representation without using
  unoccupied states},}\ }\href {\doibase 10.1021/acs.jctc.8b00123} {\bibfield
  {journal} {\bibinfo  {journal} {J. Chem. Theory Comput.}\ }\textbf {\bibinfo
  {volume} {14}},\ \bibinfo {pages} {3656--3666} (\bibinfo {year}
  {2018})}\BibitemShut {NoStop}%
\bibitem [{\citenamefont {Kullback}(1968)}]{Kullback1968}%
  \BibitemOpen
  \bibfield  {author} {\bibinfo {author} {\bibfnamefont {S.}~\bibnamefont
  {Kullback}},\ }\href@noop {} {\emph {\bibinfo {title} {Information Theory and
  Statistics}}}\ (\bibinfo  {publisher} {Dover Publications},\ \bibinfo {year}
  {1968})\BibitemShut {NoStop}%
\bibitem [{\citenamefont {Kalos}\ and\ \citenamefont
  {Whitlock}(2009)}]{kalos2009monte}%
  \BibitemOpen
  \bibfield  {author} {\bibinfo {author} {\bibfnamefont {M.}~\bibnamefont
  {Kalos}}\ and\ \bibinfo {author} {\bibfnamefont {P.}~\bibnamefont
  {Whitlock}},\ }\href@noop {} {\emph {\bibinfo {title} {Monte Carlo
  Methods}}}\ (\bibinfo  {publisher} {Wiley},\ \bibinfo {year}
  {2009})\BibitemShut {NoStop}%
\bibitem [{\citenamefont {Bayne}\ and\ \citenamefont
  {Chakraborty}(2018)}]{bayne2018development}%
  \BibitemOpen
  \bibfield  {author} {\bibinfo {author} {\bibfnamefont {M.~G.}\ \bibnamefont
  {Bayne}}\ and\ \bibinfo {author} {\bibfnamefont {A.}~\bibnamefont
  {Chakraborty}},\ }\bibfield  {title} {\enquote {\bibinfo {title} {Development
  of composite control-variate stratified sampling approach for efficient
  stochastic calculation of molecular integrals},}\ }\href@noop {} {\
  (\bibinfo {year} {2018})},\ \Eprint {http://arxiv.org/abs/1804.01197}
  {arXiv:1804.01197 [physics.chem-ph]} \BibitemShut {NoStop}%
\bibitem [{\citenamefont {Seritan}\ \emph {et~al.}(2020)\citenamefont
  {Seritan}, \citenamefont {Bannwarth}, \citenamefont {Fales}, \citenamefont
  {Hohenstein}, \citenamefont {Kokkila-Schumacher}, \citenamefont {Luehr},
  \citenamefont {Snyder}, \citenamefont {Song}, \citenamefont {Titov},
  \citenamefont {Ufimtsev},\ and\ \citenamefont
  {Martínez}}]{Seritan2020224110}%
  \BibitemOpen
  \bibfield  {author} {\bibinfo {author} {\bibfnamefont {S.}~\bibnamefont
  {Seritan}}, \bibinfo {author} {\bibfnamefont {C.}~\bibnamefont {Bannwarth}},
  \bibinfo {author} {\bibfnamefont {B.}~\bibnamefont {Fales}}, \bibinfo
  {author} {\bibfnamefont {E.}~\bibnamefont {Hohenstein}}, \bibinfo {author}
  {\bibfnamefont {S.}~\bibnamefont {Kokkila-Schumacher}}, \bibinfo {author}
  {\bibfnamefont {N.}~\bibnamefont {Luehr}}, \bibinfo {author} {\bibfnamefont
  {J.}~\bibnamefont {Snyder}, \bibfnamefont {J.W.}}, \bibinfo {author}
  {\bibfnamefont {C.}~\bibnamefont {Song}}, \bibinfo {author} {\bibfnamefont
  {A.}~\bibnamefont {Titov}}, \bibinfo {author} {\bibfnamefont
  {I.}~\bibnamefont {Ufimtsev}}, \ and\ \bibinfo {author} {\bibfnamefont
  {T.}~\bibnamefont {Martínez}},\ }\bibfield  {title} {\enquote {\bibinfo
  {title} {Terachem: Accelerating electronic structure and ab initio molecular
  dynamics with graphical processing units},}\ }\href {\doibase
  10.1063/5.0007615} {\bibfield  {journal} {\bibinfo  {journal} {The Journal of
  chemical physics}\ }\textbf {\bibinfo {volume} {152}},\ \bibinfo {pages}
  {224110} (\bibinfo {year} {2020})}\BibitemShut {NoStop}%
\bibitem [{\citenamefont {Shao}\ \emph {et~al.}(2015)\citenamefont {Shao},
  \citenamefont {Gan}, \citenamefont {Epifanovsky}, \citenamefont {Gilbert},
  \citenamefont {Wormit}, \citenamefont {Kussmann}, \citenamefont {Lange},
  \citenamefont {A.Behn}, \citenamefont {Deng},\ and\ \citenamefont
  {\textit{et. al}}}]{QCHEM_shortlist}%
  \BibitemOpen
  \bibfield  {author} {\bibinfo {author} {\bibfnamefont {Y.}~\bibnamefont
  {Shao}}, \bibinfo {author} {\bibfnamefont {Z.}~\bibnamefont {Gan}}, \bibinfo
  {author} {\bibfnamefont {E.}~\bibnamefont {Epifanovsky}}, \bibinfo {author}
  {\bibfnamefont {A.}~\bibnamefont {Gilbert}}, \bibinfo {author} {\bibfnamefont
  {M.}~\bibnamefont {Wormit}}, \bibinfo {author} {\bibfnamefont
  {J.}~\bibnamefont {Kussmann}}, \bibinfo {author} {\bibfnamefont
  {A.}~\bibnamefont {Lange}}, \bibinfo {author} {\bibnamefont {A.Behn}},
  \bibinfo {author} {\bibfnamefont {J.}~\bibnamefont {Deng}}, \ and\ \bibinfo
  {author} {\bibnamefont {\textit{et. al}}},\ }\bibfield  {title} {\enquote
  {\bibinfo {title} {Advances in molecular quantum chemistry contained in the
  q-chem 4 program package},}\ }\href {\doibase 10.1080/00268976.2014.952696}
  {\bibfield  {journal} {\bibinfo  {journal} {Molecular Physics}\ }\textbf
  {\bibinfo {volume} {113}},\ \bibinfo {pages} {184--215} (\bibinfo {year}
  {2015})}\BibitemShut {NoStop}%
\bibitem [{\citenamefont {Guidez}\ and\ \citenamefont
  {Aikens}(2012)}]{Guidez20124190}%
  \BibitemOpen
  \bibfield  {author} {\bibinfo {author} {\bibfnamefont {E.}~\bibnamefont
  {Guidez}}\ and\ \bibinfo {author} {\bibfnamefont {C.}~\bibnamefont
  {Aikens}},\ }\bibfield  {title} {\enquote {\bibinfo {title} {Theoretical
  analysis of the optical excitation spectra of silver and gold nanowires},}\
  }\href {\doibase 10.1039/c2nr30253e} {\bibfield  {journal} {\bibinfo
  {journal} {Nanoscale}\ }\textbf {\bibinfo {volume} {4}},\ \bibinfo {pages}
  {4190--4198} (\bibinfo {year} {2012})}\BibitemShut {NoStop}%
\end{thebibliography}%
\end{document}